\input harvmac.tex
%\draftmode 
%%                              TABLEAUX.TEX
%%      This  macro file is for producing a ``Young Tableau'' which is
%%      an array of little squares sometimes used in mathematical physics.
%%      For instance, the command $\tableau{6 3 2}$ will produce a tableau
%%      with 6 squares in the top row, 3 in the next, and 2 in the last.
%%                                  OOOOOO
%%      This tableau will look like OOO    but made of squares instead of O's.
%%                                  OO
%%      Any number of rows may be present, each having a nonzero number of
%%      squares.
%%
%%      A tableau is math mode material, so use $ or $$ to enclose it.
%%
%%      The size and line-thickness of the little boxes are controlled by the
%%      dimension parameters --
%%              \tableauside=1.0ex              %(size)
%%              \tableaurule=0.4pt              %(line-thickness)
%%      Change them if you want.
%%
%%                                                      -- Doug Eardley 9/19/8%%
%%
\newdimen\tableauside\tableauside=1.0ex
\newdimen\tableaurule\tableaurule=0.4pt
\newdimen\tableaustep
\def\phantomhrule#1{\hbox{\vbox to0pt{\hrule height\tableaurule width#1\vss}}}
\def\phantomvrule#1{\vbox{\hbox to0pt{\vrule width\tableaurule height#1\hss}}}
\def\sqr{\vbox{%
  \phantomhrule\tableaustep
  \hbox{\phantomvrule\tableaustep\kern\tableaustep\phantomvrule\tableaustep}%
  \hbox{\vbox{\phantomhrule\tableauside}\kern-\tableaurule}}}
\def\squares#1{\hbox{\count0=#1\noindent\loop\sqr
  \advance\count0 by-1 \ifnum\count0>0\repeat}}
\def\tableau#1{\vcenter{\offinterlineskip
  \tableaustep=\tableauside\advance\tableaustep by-\tableaurule
  \kern\normallineskip\hbox
    {\kern\normallineskip\vbox
      {\gettableau#1 0 }%
     \kern\normallineskip\kern\tableaurule}% 
  \kern\normallineskip\kern\tableaurule}}
\def\gettableau#1 {\ifnum#1=0\let\next=\null\else
  \squares{#1}\let\next=\gettableau\fi\next}

\tableauside=1.0ex
\tableaurule=0.4pt
\input epsf
\noblackbox

\def\ra{{\rightarrow}}

\def\l{{\lambda}}

%\def\subsubsec#1{$\underline{\rm #1}$}

% Something to deal with sub-sub-sections

\def\unlockat{\catcode`\@=11}
\def\lockat{\catcode`\@=12}

\unlockat
% Something to deal with sub-sub-sections

\def\newsec#1{\global\advance\secno by1\message{(\the\secno. #1)}
\global\subsecno=0\global\subsubsecno=0\eqnres@t\noindent
{\bf\the\secno. #1}
\writetoca{{\secsym} {#1}}\par\nobreak\medskip\nobreak}
%r
\global\newcount\subsecno \global\subsecno=0
\def\subsec#1{\global\advance\subsecno
by1\message{(\secsym\the\subsecno. #1)}
\ifnum\lastpenalty>9000\else\bigbreak\fi\global\subsubsecno=0
\noindent{\it\secsym\the\subsecno. #1}
\writetoca{\string\quad {\secsym\the\subsecno.} {#1}}
\par\nobreak\medskip\nobreak}
\global\newcount\subsubsecno \global\subsubsecno=0
\def\subsubsec#1{\global\advance\subsubsecno
\message{(\secsym\the\subsecno.\the\subsubsecno. #1)}
\ifnum\lastpenalty>9000\else\bigbreak\fi
\noindent\quad{\secsym\the\subsecno.\the\subsubsecno.}{#1}
\writetoca{\string\qquad{\secsym\the\subsecno.\the\subsubsecno.}{#1}}
\par\nobreak\medskip\nobreak}

\def\subsubseclab#1{\DefWarn#1\xdef
#1{\noexpand\hyperref{}{subsubsection}%
{\secsym\the\subsecno.\the\subsubsecno}%
{\secsym\the\subsecno.\the\subsubsecno}}%
\writedef{#1\leftbracket#1}\wrlabeL{#1=#1}}% Macros for boxes
\lockat

\def\IL{{\relax{\rm I\kern-.18em L}}}
\def\IH{{\relax{\rm I\kern-.18em H}}}
\def\IR{{\relax{\rm I\kern-.18em R}}}
\def\IE{{\relax{\rm I\kern-.18em E}}}
\def\IC{{\relax\hbox{$\inbar\kern-.3em{\rm C}$}}}
\def\IZ{{\relax\ifmmode\mathchoice
{\hbox{\cmss Z\kern-.4em Z}}{\hbox{\cmss Z\kern-.4em Z}}
{\lower.9pt\hbox{\cmsss Z\kern-.4em Z}}
{\lower1.2pt\hbox{\cmsss Z\kern-.4em Z}}\else{\cmss Z\kern-.4em
Z}\fi}}

\def\CP {{\cal P }}

\def\CO {{\cal O}}

\def\CS {{\cal S}}

%% MORE MACROS

\def\CO {{\cal O}}

\def\CP {{\cal P }}

\def\CS {{\cal S }}
\def\CT{{\cal T}}

\def\wb {\bar{w}}
\font\manual=manfnt \def\dbend{\lower3.5pt\hbox{\manual\char127}}

\def\IZ{{\relax\ifmmode\mathchoice
{\hbox{\cmss Z\kern-.4em Z}}{\hbox{\cmss Z\kern-.4em Z}}
{\lower.9pt\hbox{\cmsss Z\kern-.4em Z}}
{\lower1.2pt\hbox{\cmsss Z\kern-.4em Z}}\else{\cmss Z\kern-.4em
Z}\fi}}

\def\wq{{\widetilde q}}
\def\wbeta{{\widetilde \beta}}

\def\CO {{\cal O}}

\def\CP {{\cal P }}

\def\CS {{\cal S }}

\def\om{{\overline M}}

% more macros, alphabetically

\def\Aut{{\rm Aut}}

\def\IZ{{\relax\ifmmode\mathchoice
{\hbox{\cmss Z\kern-.4em Z}}{\hbox{\cmss Z\kern-.4em Z}}
{\lower.9pt\hbox{\cmsss Z\kern-.4em Z}}
{\lower1.2pt\hbox{\cmsss Z\kern-.4em Z}}\else{\cmss Z\kern-.4em
Z}\fi}}
\def\IB{{\relax{\rm I\kern-.18em B}}}
\def\IC{{\relax\hbox{$\inbar\kern-.3em{\rm C}$}}}
\def\ID{{\relax{\rm I\kern-.18em D}}}
\def\IE{{\relax{\rm I\kern-.18em E}}}
\def\IF{{\relax{\rm I\kern-.18em F}}}
\def\IG{{\relax\hbox{$\inbar\kern-.3em{\rm G}$}}}
\def\IGa{{\relax\hbox{${\rm I}\kern-.18em\Gamma$}}}
\def\IH{{\relax{\rm I\kern-.18em H}}}
\def\II{{\relax{\rm I\kern-.18em I}}}
\def\IK{{\relax{\rm I\kern-.18em K}}}
\def\IP{{\relax{\rm I\kern-.18em P}}}

\def\l{{\lambda}}

\def\inbar{\,\vrule height1.5ex width.4pt depth0pt}

\font\cmss=cmss10 \font\cmsss=cmss10 at 7pt
\def\IR{\relax{\rm I\kern-.18em R}}
\def\IT{\relax{\rm I\kern-.45em T}}

\def\Tr{{\rm Tr}}

\def\wb{{\bar{w}}}

% Macros for boxes

\def\boxit#1{\vbox{\hrule\hbox{\vrule\kern8pt
\vbox{\hbox{\kern8pt}\hbox{\vbox{#1}}\hbox{\kern8pt}}
\kern8pt\vrule}\hrule}}
\def\mathboxit#1{\vbox{\hrule\hbox{\vrule\kern8pt\vbox{\kern8pt
\hbox{$\displaystyle #1$}\kern8pt}\kern8pt\vrule}\hrule}}

%% ANOTHER SET OF MACROS

\def\inbar{\,\vrule height1.5ex width.4pt depth0pt}

\font\cmss=cmss10 \font\cmsss=cmss10 at 7pt
\def\IR{\relax{\rm I\kern-.18em R}}

\def\Tr{{\rm Tr}}

\def\wb{{\bar{w}}}

%%%%%%%%%%%%%%%%%%%%%%%%%%%%%%%%%%%%%%%%%%%%%%%%%%%%

\def\ra{{\longrightarrow}}

\def\r{{\rangle}}

%%%%%%%%%%%%%%%%%%%%%%%%%%%%%%%%%%%%%%%%%%%%%%%%%%%%%%%
\let\includefigures=\iftrue
\newfam\black
\includefigures

\input epsf
\def\plb#1 #2 {Phys. Lett. {\bf B#1} #2 }
\long\def\del#1\enddel{}
\long\def\new#1\endnew{{\bf #1}}
\let\<\langle \let\>\rangle

\def\figin{\epsfcheck\figin}\def\figins{\epsfcheck\figins}
\def\epsfcheck{\ifx\epsfbox\UnDeFiNeD
\message{(NO epsf.tex, FIGURES WILL BE IGNORED)}
\gdef\figin##1{\vskip2in}\gdef\figins##1{\hskip.5in} blank space instead
\else\message{(FIGURES WILL BE INCLUDED)}
\gdef\figin##1{##1}\gdef\figins##1{##1}\fi}
\def\DefWarn#1{}
\def\figinsert{\goodbreak\midinsert}
\def\ifig#1#2#3{\DefWarn#1\xdef#1{fig.~\the\figno}
\writedef{#1\leftbracket fig.\noexpand~\the\figno}
\figinsert\figin{\centerline{#3}}\medskip
\centerline{\vbox{\baselineskip12pt
\advance\hsize by -1truein\noindent
\footnotefont{\bf Fig.~\the\figno:} #2}}
\bigskip\endinsert\global\advance\figno by1}
%%%
\else
\def\ifig#1#2#3{\xdef#1{fig.~\the\figno}
\writedef{#1\leftbracket fig.\noexpand~\the\figno}
\figinsert\figin{\centerline{#3}}\medskip
\centerline{\vbox{\baselineskip12pt
\advance\hsize by -1truein\noindent
\footnotefont{\bf Fig.~\the\figno:} #2}}
\bigskip\endinsert
\global\advance\figno by1}
\fi

\input xy
\xyoption{all}
\font\cmss=cmss10 \font\cmsss=cmss10 at 7pt
\def\inbar{\,\vrule height1.5ex width.4pt depth0pt}
\def\IC{{\relax\hbox{$\inbar\kern-.3em{\rm C}$}}}
\def\IP{{\relax{\rm I\kern-.18em P}}}
\def\IF{{\relax{\rm I\kern-.18em F}}}
\def\IZ{\relax\ifmmode\mathchoice
{\hbox{\cmss Z\kern-.4em Z}}{\hbox{\cmss Z\kern-.4em Z}}
{\lower.9pt\hbox{\cmsss Z\kern-.4em Z}}
{\lower1.2pt\hbox{\cmsss Z\kern-.4em Z}}\else{\cmss Z\kern-.4em
Z}\fi}
\def\IR{{\relax{\rm I\kern-.18em R}}}
\def\IQ{\relax\hbox{\kern.25em$\inbar\kern-.3em{\rm Q}$}}

%%% special math symbols
\def\pmb#1{\setbox0=\hbox{#1}%
 \kern-.025em\copy0\kern-\wd0
 \kern.05em\copy0\kern-\wd0
 \kern-.025em\raise.0433em\box0 }
\font\cmss=cmss10
\font\cmsss=cmss10 at 7pt
\def\rlx{\relax\leavevmode}
\def\Cop{\relax\,\hbox{$\inbar\kern-.3em{\rm C}$}}
\def\Rop{\relax{\rm I\kern-.18em R}}
\def\Nop{\relax{\rm I\kern-.18em N}}
\def\Pop{\relax{\rm I\kern-.18em P}}
\def\Zop{\rlx\leavevmode\ifmmode\mathchoice{\hbox{\cmss Z\kern-.4em Z}}
 {\hbox{\cmss Z\kern-.4em Z}}{\lower.9pt\hbox{\cmsss Z\kern-.36em Z}}
 {\lower1.2pt\hbox{\cmsss Z\kern-.36em Z}}\else{\cmss Z\kern-.4em
 Z}\fi}

\def\inbar{\,\vrule height1.5ex width.4pt depth0pt}
\def\IC{{\relax\hbox{$\inbar\kern-.3em{\rm C}$}}}
\def\IP{{\relax{\rm I\kern-.18em P}}}
\def\IF{{\relax{\rm I\kern-.18em F}}}
\def\IZ{\relax\ifmmode\mathchoice
{\hbox{\cmss Z\kern-.4em Z}}{\hbox{\cmss Z\kern-.4em Z}}
{\lower.9pt\hbox{\cmsss Z\kern-.4em Z}}
{\lower1.2pt\hbox{\cmsss Z\kern-.4em Z}}\else{\cmss Z\kern-.4em
Z}\fi}
\def\IR{{\relax{\rm I\kern-.18em R}}}
\def\IT{{\mathchoice {\setbox0=\hbox{$\displaystyle\rm
T$}\hbox{\hbox to0pt{\kern0.3\wd0\vrule height0.9\ht0\hss}\box0}}
{\setbox0=\hbox{$\textstyle\rm T$}\hbox{\hbox
to0pt{\kern0.3\wd0\vrule height0.9\ht0\hss}\box0}}
{\setbox0=\hbox{$\scriptstyle\rm T$}\hbox{\hbox
to0pt{\kern0.3\wd0\vrule height0.9\ht0\hss}\box0}}
{\setbox0=\hbox{$\scriptscriptstyle\rm T$}\hbox{\hbox
to0pt{\kern0.3\wd0\vrule height0.9\ht0\hss}\box0}}}}
\def\bbbti{{\mathchoice {\setbox0=\hbox{$\displaystyle\rm
T$}\hbox{\hbox to0pt{\kern0.3\wd0\vrule height0.9\ht0\hss}\box0}}
{\setbox0=\hbox{$\textstyle\rm T$}\hbox{\hbox
to0pt{\kern0.3\wd0\vrule height0.9\ht0\hss}\box0}}
{\setbox0=\hbox{$\scriptstyle\rm T$}\hbox{\hbox
to0pt{\kern0.3\wd0\vrule height0.9\ht0\hss}\box0}}
{\setbox0=\hbox{$\scriptscriptstyle\rm T$}\hbox{\hbox
to0pt{\kern0.3\wd0\vrule height0.9\ht0\hss}\box0}}}}

\def\F{{\cal{F}}}

\def\I{{\cal I}}

\def\ox{{\overline X}}
\def\bx{{\overline x}}
\def\by{{\overline y}}

\def\r{{\rho}}
\def\Def{{\rm Def}}
%%%%%%%%%%%%%%%%%%%%%%%%%%%%%%%%%%%%%%%%%%%%%%%%%%%%%%%%%%%%%%%%%%%%%%
%\nref\AKV{M. Aganagic, A. Klemm and C. Vafa, ``Disk Instantons,
%Mirror Symmetry and the Duality Web'', Z. Naturforsch. {\bf A 57}
%(2002) 1, hep-th/0105045.}
\nref\AMV{M. Aganagic, M. Mari{\~n}o and C. Vafa, 
``All Loop Topological String Amplitudes from Chern-Simons 
Theory'', hep-th/0206164.}
\nref\AKMV{M. Aganagic, A. Klemm, M. Marino and C. Vafa, ``The Topological 
Vertex'', hep-th/0305132.}
\nref\BF{K. Behrend and B. Fantechi, ``The Intrinsic Normal Cone'',
Invent. Math. {\bf 128} (1997) 45.}
%\nref\CKYZ{T.-M. Chiang, A. Klemm, S.-T. Yau and E. Zaslow, ``Local Mirror Symmetry: Calculations and
%Interpretations'', ATMP {\bf 3} (1999) 495, hep-th/9903053.}
\nref\DFGi{D.-E. Diaconescu, B. Florea and A. Grassi,
``Geometric Transitions and Open String Instantons'', ATMP {\bf 6} (2002) 619, hep-th/0205234.}
\nref\DFGii{D.-E. Diaconescu, B. Florea and A. Grassi, 
``Geometric Transitions, del Pezzo Surfaces and Open String Instantons'', ATMP {\bf 6} (2002) 643, 
hep-th/0206163.} 
\nref\DF{D.-E. Diaconescu and B. Florea, ``Large $N$ Duality for Compact Calabi-Yau Threefolds'', 
hep-th/0302076.}
\nref\F{C. Faber, ``Algorithms for Computing Intersection Numbers on Moduli Spaces of Curves, with an 
Application to the Class of the Locus of Jacobians'', in {\it New Trends in Algebraic Geometry}, Cambridge Univ. Press., 1999, 
alg-geom/9706006.}
\nref\GViii{R. Gopakumar and C. Vafa, ``On the Gauge Theory/Geometry
Correspondence'', ATMP {\bf 3} (1999) 1415, hep-th/9811131.}
\nref\GP{T. Graber and R. Pandharipande,
``Localization of Virtual Classes'', Invent. Math. {\bf 135}
(1999) 487, math.AG/9708001.}
%\nref\GVi{R. Gopakumar and C. Vafa, ``Topological Gravity as Large
%$N$ Topological Gauge Theory'', ATMP {\bf 2} (1998) 413,
%hep-th/9802016.}
%\nref\GVii{R. Gopakumar and C. Vafa, `` M-Theory and Topological
%Strings -- I'', hep-th/9809187;  `` M-Theory and Topological Strings -- II'', hep-th/9812127.}
\nref\GZ{T. Graber and E. Zaslow, ``Open String Gromov-Witten
Invariants: Calculations and a Mirror 'Theorem' '', hep-th/0109075.}
%\nref\MG{M. Gross, ``Topological Mirror Symmetry'',
%math.AG/9909015.}
\nref\I{A. Iqbal, ``All Genus Topological String Amplitudes and 5-brane Webs as 
Feynman Diagrams'', hep-th/0207114.}
\nref\IK{A. Iqbal and A.-K. Kashani-Poor, ``$SU(N)$ Geometries and Topological 
String Amplitudes'', hep-th/0306032.}
%\nref\KKLMi{S.
%Kachru, S. Katz, A. Lawrence and J. McGreevy, ``Open String
%Instantons and Superpotentials'', Phys. Rev. {\bf D62} (2000)
%026001, hep-th/9912151.}
%\nref\KKLMii{S. Kachru, S. Katz, A.
%Lawrence and J. McGreevy, ``Mirror Symmetry for Open Strings'',
%hep-th/0006047.}
\nref\KL{S. Katz and C.-C. M. Liu, ``Enumerative Geometry of Stable
Maps with Lagrangian Boundary Conditions and Multiple Covers of
the Disc'', ATMP {\bf 5} (2001) 1, math.AG/0103074.}
\nref\MK{M. Kontsevich, ``Enumeration of Rational Curves via Torus Actions'',
in {\it The Moduli Space of Curves}, 335-368, Progr. Math. {\bf 129},
Birkh\"auser Boston, MA, 1995.}
\nref\LMV{J.M.F. Labastida, M. Mari\~no and C. Vafa, ``Knots,
Links and Branes at Large $N$'', JHEP {\bf 11} (2000) 007, hep-th/0010102.}
\nref\LT{J. Li and G. Tian,
``Virtual Moduli Cycles and Gromov-Witten Invariants of Algebraic Varieties'',
J. Amer. Math. Soc. {\bf 11} (1998) 119.}
\nref\Lii{J. Li, ``A Degeneration of Stable Morphisms and Relative Stable Morphisms'', 
math.AG/0009097.}
\nref\Liii{J. Li, ``A Degeneration Formula of GW-invariants'', math.AG/0110113.}
\nref\LS{J. Li and Y.S. Song, ``Open String Instantons and Relative Stable
Morphisms'', ATMP {\bf 5} (2002) 67, hep-th/0103100.}
\nref\ML{C.-C. M. Liu, ``Moduli of $J$-Holomorphic Curves with Lagrangian Boundary Conditions and 
Open Gromov-Witten Invariants for an $S^1$-Equivariant Pair'', math.SG/0210257.}
\nref\LLZi{C.-C. M. Liu, K. Liu and J. Zhou, ``On a Proof of a Conjecture of Mari\~no-Vafa on 
Hodge Integrals'', math.AG/0306257.}
\nref\LLZii{C.-C. M. Liu, K. Liu and J. Zhou, ``A Proof of a Conjecture of Mari\~no-Vafa on 
Hodge Integrals'', math.AG/0306434.}
%\nref\LMi{J.M.F. Labastida and M. Mari\~no, ``Polynomial Invariants for
%Torus Knots and Topological Strings'',  Commun. Math. Phys. {\bf 217} (2001)
%423, hep-th/0004196.}
%\nref\LMV{J.M.F. Labastida, M. Mari\~no and C. Vafa, ``Knots,
%Links and Branes at Large $N$'', JHEP {\bf 11} (2000) 007, hep-th/0010102.}
%\nref\LMii{J.M.F. Labastida and M. Mari\~no, ``A New Point of View in the
%Theory of Knot and Link Invariants'', math.QA/0104180.}
\nref\MV{M. Mari\~no and C. Vafa, ``Framed Knots at Large $N$'',
hep-th/0108064.}
\nref\Mii{P. Mayr, ``Summing up Open String Instantons and ${\cal
N}=1$ String Amplitudes'', hep-th/0203237.}
\nref\OP{A. Okounkov and R. Pandharipande, ``Hodge Integrals and Invariants of the Unknot'', 
math.AG/0307209.}
\nref\OV{H. Ooguri and C. Vafa, ``Knot Invariants and Topological
Strings'', Nucl. Phys. {\bf B 577} (2000) 419, hep-th/9912123.}
%\nref\OVii{H. Ooguri and C. Vafa, 
%``Worldsheet Derivation of a Large $N$ Duality'', hep-th/0205297.}
%\nref\RS{P. Ramadevi and T. Sarkar,
%``On Link Invariants and Topological String Amplitudes'',
%Nucl. Phys. {\bf B600} (2001) 487, hep-th/0009188.}
%\nref\EWii{E.
%Witten, ``Chern-Simons Gauge Theory as a String Theory'',
%``The Floer Memorial Volume'', H. Hofer, C.H. Taubes, A. Weinstein
%and E. Zehnder, eds, Birkh\"auser 1995, 637,
%hep-th/9207094.}
%%%%%%%%%%%%%%%%%%%%%%%%%%%%%%%%%%%%%%%%%%%%%%%%%%%%%%%%%%%%%%%%%%%%
\Title{
\vbox{
\baselineskip12pt
\hbox{hep-th/0309143}
\hbox{RUNHETC-2003-27}}}
{\vbox{\vskip 37pt
\vbox{\centerline{Localization and Gluing of Topological Amplitudes}}
}}
\vskip 15pt
\centerline{Duiliu-Emanuel Diaconescu\footnote{$^\natural$}{{{\tt duiliu@physics.rutgers.edu}}} 
and Bogdan Florea\footnote{$^\sharp$}{{{\tt florea@physics.rutgers.edu}}}} 
\bigskip
\medskip
\centerline{{\it Department of Physics and Astronomy,
Rutgers University,}}
\centerline{\it Piscataway, NJ 08855-0849, USA}
\bigskip
\bigskip
\bigskip
\bigskip
\smallskip
\noindent 
We develop a gluing algorithm for Gromov-Witten invariants of toric Calabi-Yau 
threefolds based on localization and gluing graphs. The main building 
block of this algorithm is a generating function of cubic Hodge 
integrals of special form. We conjecture a precise relation between this generating function and 
the topological vertex at fractional framing.  

\vfill
\Date{September 2003}

\newsec{Introduction}

A gluing algorithm for topological amplitudes on toric Calabi-Yau threefolds has been 
recently constructed in \AKMV.\ This algorithm is based on gluing topological vertices 
derived from large $N$ duality and Chern-Simons theory. Previous work on the subject 
can be found in \refs{\AMV,\DFGi,\DFGii,\DF,\I,\IK}. In this paper we develop a parallel
enumerative algorithm relying on localization and gluing of graphs. The main building block 
of this construction is a generating functional of cubic Hodge integrals, which is related to 
the topological vertex of \AKMV.\

The paper is structured as follows. Section two is a review of local Gromov-Witten invariants 
associated to noncompact toric threefolds, localization and graphs. In section three we develop 
an algorithm for cutting and pasting of graphs from a pure combinatoric point of view. 
A concrete geometric implementation of this algorithm is presented in section four. 
The unit block can be formally written as a topological open string partition function 
for three lagrangian cycles in $\IC^3$. Applying open string localization \refs{\GZ,\KL,\LS}, 
we obtain a generating function for cubic Hodge integrals.
Section five is devoted to a detailed comparison of this function with
the topological vertex of \AKMV.\  
We conjecture that the two expressions agree 
provided that the topological vertex is evaluated at fractional framing. 
A special case of this conjecture corresponding to a vertex with two trivial representations 
has been recently proved in \refs{\LLZi,\LLZii,\OP}.
We present strong numerical evidence for the general case by direct computations, but the 
proof is an open problem. 
Some technical details are included in two appendixes. 

{\it Acknowledgments.} We are very grateful to Antonella Grassi for collaboration at an early 
stage of this project, especially for invaluable help with the graph combinatorics in section three. 
We would also like to thank Mina Aganagic, Sheldon Katz, Amir Kashani-Poor, Melissa Liu, 
Marcos Mari{\~ n}o and 
Cumrun Vafa for helpful discussions and Carel Faber for kindly sending us the Maple 
implementation of 
the algorithm \F.\ The work of D.-E.D. has been partially supported by DOE grant 
DOE-DE-FG02-96ER40959 
and by an Alfred P. Sloan foundation fellowship. The work of B.F. 
has been partially supported by DOE grant DOE-DE-FG02-96ER40959. We would also like to ackenowledge 
the hospitality of KITP Santa Barbara where part of this work has been done.

\newsec{Localization and Graphs} 

Let $X$ be a smooth projective Calabi-Yau threefold. The Gromov-Witten invariants of 
$X$ are defined in terms of intersection theory on the moduli space of stable 
maps $\om_{g,0}(X,\beta)$ with fixed homology class $\beta \in H_2(X)$. More specifically, 
the moduli space $\om_{g,0}(X,\beta)$ has a special structure -- perfect obstruction theory -- 
which produces a virtual fundamental cycle of expected dimension 
$[\om_{g,0}(X,\beta)]^{vir}\in A_0(\om_{g,0}(X,\beta))$.
One defines the Gromov-Witten potential $F_X(g_s,q)$ as a formal series 
\eqn\gwa{
F_X(g_s,q) =\sum_{g=0}^\infty \sum_{\beta\in H_2(X)} C_{g,\beta} q^\beta} 
where 
\eqn\gwb{
C_{g,\beta} = \int_{[\om_{g,0}(X,\beta)]^{vir}} 1.}
Here $q^\beta$ is a formal multisymbol satisfying $q^{\beta+\beta'}=q^\beta q^{\beta'}$.

\subsec{Local Gromov-Witten Invariants} 

In this paper we are mainly interested in noncompact toric Calabi-Yau threefolds $X$. 
The previous definition has to be refined since the moduli space $\om_{g,0}(X,\beta)$ 
is in principle ill-defined. 
Let $\ox$ be a projective completion of $X$ 
so that the divisor at infinity $D=\ox \setminus X$ is reduced with normal crossings.
Then there is a well-defined moduli space $\om_{g,0}(\ox,D,\beta)$ of relative stable 
maps to the pair $(\ox,D)$ with multiplicity zero along $D$. 
Moreover, this moduli space has a well-defined perfect obstruction theory and a 
virtual fundamental cycle $[\om_{g,0}(\ox,D,\beta)]^{vir}$ \refs{\Lii,\Liii}.
For a class $\beta \in H_2(X)$, 
this moduli space may contain closed connected components parameterizing maps 
supported away form $D$. We will denote the union of all these components by $\om_{g,0}(X,\beta)$.
The virtual cycle $[\om_{g,0}(\ox,D,\beta)]^{vir}$ induces a virtual cycle of expected dimension 
on  $\om_{g,0}(X,\beta)$ by functoriality. Therefore we can define local Gromov-Witten invariants 
as in the compact case, taking into account the new meaning of  $\om_{g,0}(X,\beta)$.
Note that $\om_{g,0}(X,\beta)$ may be empty, in which case $C_{g,\beta}=0$. 

To clarify this definition, let us consider some examples. First take $X$ to be the total 
space of the canonical bundle $K_S$ over a toric Fano surface $S$, and let $\beta$ be a 
curve class in the zero section. 
The completion can be taken to be $\ox \simeq \IP(\CO_S\oplus \CO_S(K_S))$. 
In this case $\om_{g,0}(X,\beta) \simeq \om_{g,0}(S,\beta)$, and we could have adopted 
this as a definition of the moduli space. However, there are more general cases when 
such a direct approach is not possible. Consider for example the threefold defined 
by the toric diagram below. 

\ifig\tordiag{A section in the toric fan of a local Calabi-Yau threefold $X$.}
{\epsfxsize2in\epsfbox{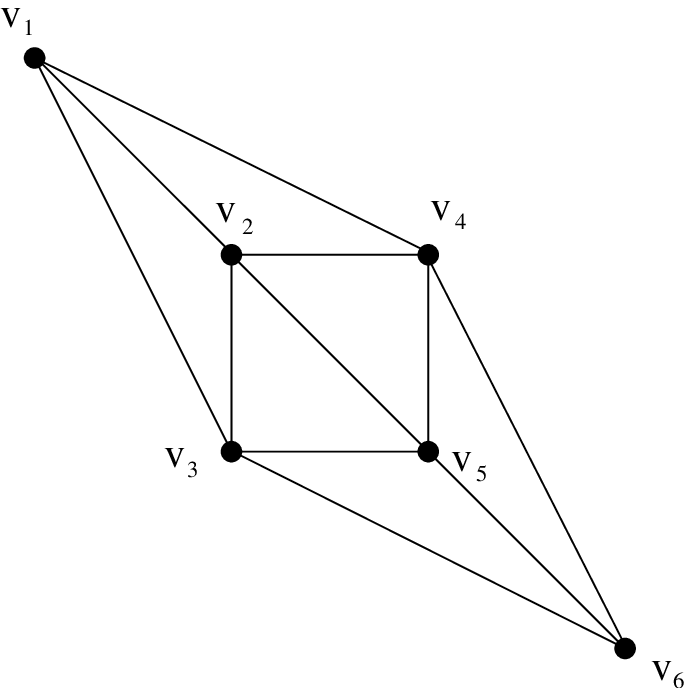}}

There are two compact divisors on $X$, $S_1, S_2$, both 
isomorphic to the Hirzebruch surface $\IF_1$, which intersect along a $(-1,-1)$ curve. 
Any curve $C$ lying on $S_1\cup S_2$ cannot be deformed in the normal directions, since 
both $S_1, S_2$ are Fano. Therefore any map $f:\Sigma \ra X$ with $f_*[\Sigma]=[C]$ is 
supported away from the divisor at infinity. One could try to define local invariants 
in terms of maps to the singular divisor $S_1\cup S_2$, but this approach would be 
quite involved. It is more convenient to use the construction explained in the 
previous paragraph, in which case the target space $X$ is smooth. 

\subsec{Localization} 

Since $X$ is toric, it admits a torus action $T\times X \ra X$ which induces an action on the 
moduli space $\om_{g,0}(X,\beta)$. Then the local Gromov-Witten invariants can be computed by 
localization \GP.\ To recall the essential aspects, the virtual cycle $[\om_{g,0}(X,\beta)]^{vir}$ 
induces a virtual cycle $[\Xi]^{vir}$ on each component $\Xi$ of the fixed locus. Moreover, 
one can 
construct a virtual normal bundle $N_\Xi^{vir}$ to each fixed locus. The localization formula 
reads 
\eqn\locA{
C_{g,\beta} =\sum_{\Xi\subset \om_{g,0}(X,\beta)} \int_{[\Xi]^{vir}} {1\over e_T(N_\Xi^{vir})}}
where $e_T$ denotes the equivariant Euler class. 

The fixed loci in the moduli space of stable maps can be indexed by graphs according to 
\MK.\ Since this construction will play an important role in the paper, let us recall 
the basic elements. Let $\{P_r\}$, $r=1,\ldots, N$ 
denote the fixed points of the torus action on $X$. 
Any two fixed points are joined by a $T$-invariant rational curve $C_{rs}$. The configuration
of invariant curves forms a graph $\Gamma$ whose vertices are in $1-1$ correspondence with 
the fixed points $P_r$ and edges in $1-1$ correspondence with curves $C_{rs}$. Note that 
at most three edges can meet at any vertex. 
Some examples are represented below. 

\ifig\graphs{The graph $\Gamma$ for $a)$ $\CO(-3)\ra \IP^2$ and $b)$ the toric Calabi-Yau 
threefold represented in fig. 1.}
{\epsfxsize2.5in\epsfbox{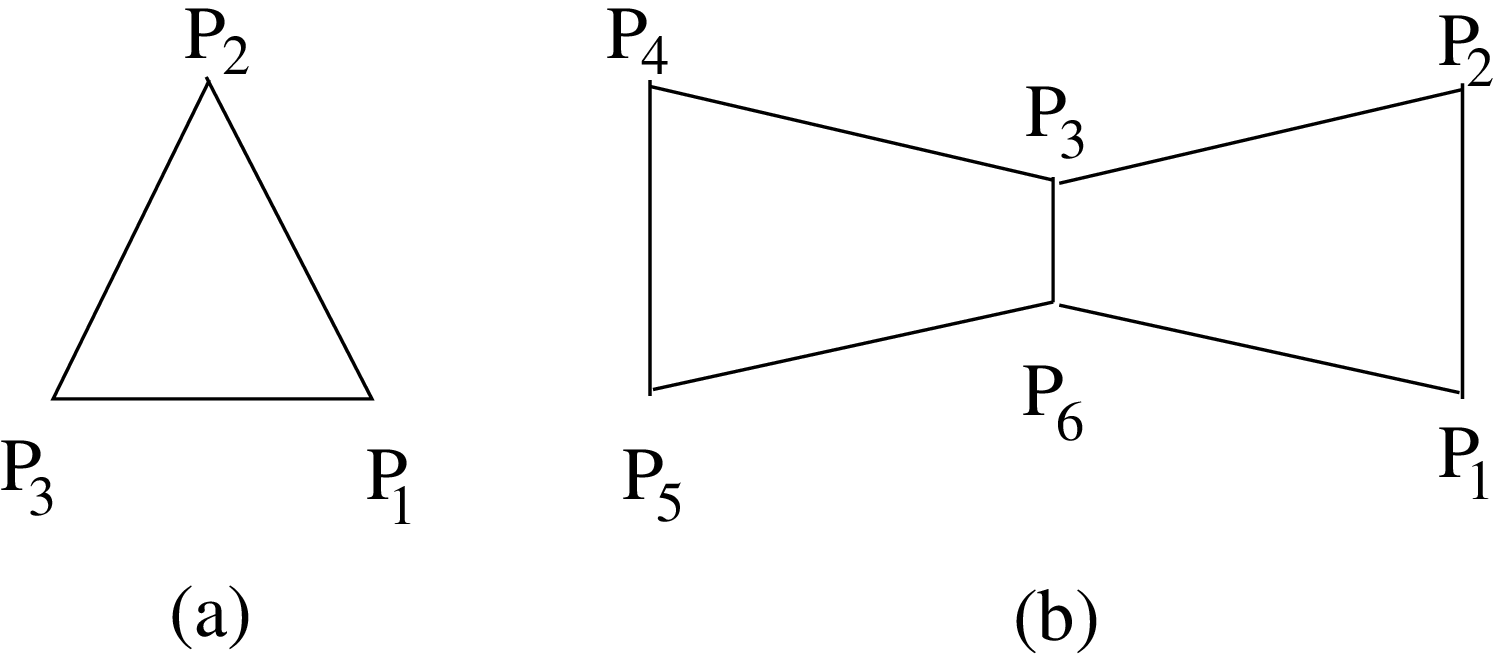}}

The fixed maps $f:\Sigma \ra X$ have a special structure. The image of $f$ is contained 
in the configuration of invariant curves $\cup_{r,s} C_{rs}$. $f^{-1}(C_{rs})$ consists of 
finitely many irreducible components of $\Sigma$ which are smooth rational curves. 
The restriction of $f$ to such a component must be a Galois cover. $f^{-1}(P_s)$ consists of 
finitely many prestable curves on $\Sigma$ possibly of higher genus. Note that higher 
genus components mapping onto some $C_{rs}$ are not allowed; all higher genus components 
must be collapsed to fixed points. 

To any irreducible component $\Xi$ of the fixed locus we can associate a connected graph 
$\Upsilon$ as follows. Let $f:\Sigma \ra X$ denote a map in $\Xi$. 

i) The vertices $v\in V(\Upsilon)$ represent 
prestable curves $\Sigma_v \subset \Sigma$ mapping to some fixed point $P_i$. 
Each vertex is marked by two numbers $(k_v,g_v)$, where $k_v\in {\{1,2,\ldots,N\}}$
is defined by $f(\Sigma_v)=P_{k_v}$, and $g_v$ is the arithmetic genus of $\Sigma_v$. 
Note that $\Sigma_v$ may be a point. 

ii) The edges $e\in E(\Upsilon)$ correspond to irreducible rational components 
of $\Sigma$ mapped onto $C_{rs}$ for some $(r,s)$. 
Each edge is marked by an integer $d_e$ representing the degree of the Galois 
cover $f|_{\Sigma_e}:\Sigma_e\ra C_{rs}$. 

Let us define a flag \MK\ to be a pair $(v,e)\in V(\Upsilon)\times E(\Upsilon)$ 
so that $v\in e$. For a given 
$v\in V(\Upsilon)$ we define the valence of $v$, $val(v)$ to be the number of 
flags $(v,e)$. We will also denote by $F(\Upsilon)$ the set of flags of $\Upsilon$
and by $F_v(\Upsilon)$ the set flags with given vertex $v$. 
Geometrically, $val(v)$ counts the number of rational components 
$\Sigma_e$ which intersect a given prestable curve $\Sigma_v$. 
Each flag determines a marked point $p_{(v,e)}\in \Sigma_v$, 
so that $(\Sigma_v, p_{(v,e)})$ is a prestable curve of genus $g_v$ with $val(v)$ 
marked points. 

According to \MK,\ the set of all fixed loci $\Xi$ is 
in one to one correspondence with (equivalence classes 
of) graphs $\Upsilon$ subject to the following conditions 

1) If $e\in E(\Upsilon)$ is an edge connecting two vertices $u, v$, 
then $k_u\neq k_v$. 

2) $1-\chi(\Upsilon) +\sum_{v\in V(\Upsilon)} g_v = g$, where $g_v$ 
is the arithmetic genus of the component $\Sigma_v$, and $\chi(\Upsilon)$ 
is the Euler characteristic of $\Upsilon$, 
$\chi(\Upsilon) = |V(\Upsilon)|-|E(\Upsilon)|$. 

3) $\sum_{e\in E(\Upsilon)} d_ef_*[\Sigma_e] = \beta$.  

4) For all $v\in V(\Upsilon)$, $(\Sigma_v, p_{(v,e)})$ is a stable marked curve.

\noindent 
Note that the last condition gives rise to some special cases, 
namely $(g_v,val(v))=(0,1), (0,2)$. 
If $(g_v, val(v))=(0,1)$, $\Sigma_v = p_{v}$ is a smooth point of $\Sigma$.
If $(g_v, val(v))=(0,2)$, $\Sigma_v = p_{v}$ is a node of $\Sigma$ 
lying at the intersection of two components $\Sigma_{e_1(v)}, \Sigma_{e_2(v)}$. 

Next, let us outline the computation of the local contribution 
$\int_{[\Xi]^{vir}} {1\over e_T(N_\Xi^{vir})}$ for a fixed component $\Xi$ 
with associated graph $\Upsilon$. Given the structure of an arbitrary fixed map, 
the fixed locus $\Xi$ is isomorphic to a quotient of 
$\prod_{v\in V(\Upsilon)} \om_{g_v,val(v)}$ by a finite group $G(\Upsilon)$. 
The finite group admits a presentation 
\eqn\locAB{
1\ra \prod_{e\in E(\Upsilon)} \IZ/{d_e}\ra G(\Upsilon) \ra \Aut(\Upsilon)\ra 1}
where $\Aut(\Upsilon)$ is the automorphism group of the graph. 

The main tool is the tangent obstruction complex of a map $f:\Sigma \ra X$, which 
encodes the local structure of the moduli space near the point $(\Sigma, f)$. We have 
\eqn\locB{
0\ra \Aut(\Sigma)\ra H^0(\Sigma, f^*T_X) \ra {\IT}^1 
\ra \Def(\Sigma) \ra H^1(\Sigma, f^*T_X) \ra {\IT}^2\ra 0}
where ${\IT}^1, {\IT}^2$ are the infinitesimal deformation and respectively obstruction 
space of a map $(\Sigma ,f)$. 
$\Aut(\Sigma), \Def(\Sigma)$ denote the infinitesimal automorphism and respectively 
deformation groups of the domain $\Sigma$. Note that if $(f,\Sigma)$ represents a point in $\Xi$, 
there is an induced $T$-action on the complex \locB.\ 

According to \GP,\ the fixed part of \locB\ under the torus action determines the 
virtual cycle $[\Xi]^{vir}$ while the moving part determines the normal bundle 
$N_\Xi^{vir}$. Moreover the induced virtual class coincides with the ordinary fundamental 
class of $\Xi$ regarded as an orbispace. The integrand ${1\over e_T(N^{vir}_{\Xi})}$ 
can be computed in terms of the graph $\Upsilon$ using the normalization exact sequence
\eqn\locD{\eqalign{
& 0 \ra f^*T_X \ra \bigoplus_{e\in E(\Gamma)} f^*_eT_X \oplus \bigoplus_{v\in 
V(\Gamma)} f^*_vT_X \ra \bigoplus_{v\in V(\Gamma)} (T_{P_{k_v}}X)^{val(v)}
\ra 0.\cr}}
Note that the terms of this sequence form sheaves over the fixed locus $\Xi$ 
which may not be in general locally free. For localization computations we only 
need the equivariant K-theory classes of these sheaves which will be denoted by 
$[\ ]$. 

The associated long exact sequence of \locD\ reads 
\eqn\locE{\eqalign{
0 & \ra H^0(\Sigma, f^*T_X) \ra 
\bigoplus_{e\in E(\Gamma)}H^0(\Sigma_e, f^*_eT_X)\oplus \bigoplus_{v\in 
V(\Gamma)} T_{P_{k_v}}X 
 \ra \bigoplus_{v\in V(\Gamma)} (T_{P_{k_v}}X)^{val(v)} \cr
& \ra H^1(\Sigma, f^*T_X)\ra \bigoplus_{e\in E(\Gamma)}H^1(\Sigma_e, f^*_eT_X)
\oplus \bigoplus_{v\in V(\Gamma)} H^1(\Sigma_v, \CO_{\Sigma_v})\otimes 
T_{P_{k_v}}X\ra 0.\cr}}
This yields 
\eqn\locEB{\eqalign{
[N^{vir}_\Xi]= & \sum_{e\in E(\Upsilon)} 
\left([H^0(\Sigma_e, f_e^*T_X)^m] - [H^1(\Sigma_e,f_e^*T_X)^m]\right)\cr
& - \sum_{v\in V(\Upsilon)} \left([H^1(\Sigma_v,\CO_{\Sigma_v})\otimes T_{P_{k_v}}X] 
+ (val(v)-1)[ T_{P_{k_v}}X]\right)\cr
& + [\Def(\Sigma)^m] -[\Aut(\Sigma)^m].\cr}}
The moving part of the automorphism group consists of 
holomorphic vector fields on the horizontal components $\Sigma_e$ 
which vanish at the nodes of $\Sigma$ lying on $\Sigma_e$. We can write  
\eqn\locF{
[\Aut(\Sigma)^m]= \sum_{e\in E(\Upsilon)} [H^0(\Sigma_e, T_{\Sigma_e})^m] -
\sum_{(v,e)\in F(\Upsilon), val(v)\geq 2} [T_{p_{(v,e)}}\Sigma_e].}
The moving infinitesimal deformations of $\Sigma$ are deformations 
of the nodes lying at least on one edge component
\eqn\locG{ \eqalign{
[\Def(\Sigma)^m] = & \sum_{(v,e)\in F(\Upsilon), (g_v, val(v))\neq (0,1), (0,2)} 
[T_{p_{(v,e)}}\Sigma_e\otimes T_{p_{(v,e)}}\Sigma_v] \cr
&+\sum_{(v,e)\in F(\Upsilon), 
(g_v, val(v))=(0,2)} [T_{p_{v}}\Sigma_{e_1(v)} \otimes T_{p_{v}}\Sigma_{e_2(v)}].
\cr}}
Collecting the facts, it follows that the local contribution of the fixed locus $\Xi$ 
can be written as 
\eqn\locH{\eqalign{
\int_{[\Xi]^{vir}} {1\over e_{T}(N^{ vir}_\Xi)} = & 
{1\over |{\rm \Aut}(\Upsilon)|\prod_{e\in E(\Upsilon)}d_e}
 \prod_{e\in E(\Upsilon)}F(e)
\prod_{v\in V(\Upsilon), (g_v, val(v)) =(0,2)}G(v)\cr 
& \times\prod_{v\in V(\Upsilon), (g_v, val(v)) \neq (0,1), (0,2)} 
\int_{(\om_{g_v,val(v)})_T} H(v)\cr}}
where 
\eqn\locI{\eqalign{ 
& F(e) = {e_T(H^1(\Sigma_e, f_e^*T_X)^m) e_T(H^0(\Sigma_e,T_{\Sigma_e})^m) \over 
e_T(H^0(\Sigma_e, f_e^*T_X)^m)} \cr
& G(v) = {e_T(T_{P_{k_v}}X)\over e_T(T_{p_v}\Sigma_{e_1(v)})e_T(T_{p_v}\Sigma_{e_2(v)})
\left(e_T(T_{p_v}\Sigma_{e_1(v)})+e_T(T_{p_v}\Sigma_{e_2(v)})\right)}\cr
& H(v) = {e_T(H^1(\Sigma_v,\CO_{\Sigma_v})\otimes T_{P_{k_v}}X)\over 
\prod_{(v,e)\in F_v(\Upsilon)} \left[e_T(T_{p_{(v,e)}}\Sigma_e) (e_T(T_{p_{(v,e)}}\Sigma_e) + 
e_T(T_{p_{(v,e)}}\Sigma_v))\right]}\cr
&\qquad\  ={e_T(\IE_v^\vee\otimes  T_{P_{k_v}}X)\over 
\prod_{(v,e)\in F_v(\Upsilon)} \left[e_T(T_{p_{(v,e)}}\Sigma_e) (e_T(T_{p_{(v,e)}}\Sigma_e) -
\psi_{p_{(v,e)}})\right]}.\cr}}
In the last equation $\IE_v$ is the Hodge bundle on the Deligne-Mumford moduli space 
$\om_{g_v,val(v)}$ and $\psi_{p_{(v,e)}}$ are Mumford classes associated to the 
marked points $\{p_{(v,e)}\}$. 

To conclude this section, note that the Gromov-Witten potential \gwa\ can be written as 
a sum over marked graphs $\Upsilon$ satisfying condition (1) above equation \locAB.\ 
For each such graph we define the genus $g(\Upsilon) = 2- \chi(\Upsilon) + 
\sum_{v\in V(\Upsilon)} g_v$ 
and the homology class $\beta(\Upsilon) = \sum_{e\in \Upsilon} 
d_e f_{e*}[\Sigma_e]$. Then we have 
\eqn\cpA{
F_X(g_s,q) = \sum_{\Upsilon} {1\over |{\rm \Aut}(\Upsilon)|\prod_{e\in E(\Upsilon)}d_e}
C(\Upsilon) g_s^{2g(\Upsilon)-2} q^{\beta(\Upsilon)}.}
Note that $F_X(g_s,q)$ depends only on the marked graph 
$\Gamma$, hence we can alternatively denote it by $F_\Gamma(g_s,q)$. 

We can further reformulate \cpA\ by noting that the data $k_v, v\in V(\Upsilon)$ 
is equivalent to a map of graphs $\phi:\Upsilon\ra \Gamma$. 
Therefore  a marked graph $\Upsilon$ can be alternatively thought as a pair 
$({\widetilde \Upsilon},\phi)$ where $({\widetilde \Upsilon})$ obtained from 
$\Upsilon$ by deleting the markings $k_v$, 
and $\phi:{\widetilde\Upsilon} \ra \Gamma$ is a map of graphs. 
Condition (1) above \locAB\ is replaced by 

(1') $\phi(u)\neq \phi(v)$ for any two distinct vertices $u,v\in V({\widetilde \Upsilon})$. 

\noindent
In the following we will use the notation $(\Upsilon, \phi)$ for such a pair.

\newsec{Gluing Algorithm -- Combinatorics} 

Our goal is to find a gluing formula for the Gromov-Witten invariants of  $X$ 
based on a decomposition of the graph $\Gamma$ into smaller units. The main idea is to construct 
suitable generating functional for each such unit so that the full Gromov-Witten potential 
can be obtained by gluing these local data. 
In this section we will discuss purely combinatoric aspects of this algorithm. 
A geometric realization will be presented in the next section. 

To review our setup, we are given a 
graph $\Gamma$ satisfying the following conditions 

$i)$ There are no edges starting and ending at the same vertex.  

$ii)$ Any two distinct vertices are joined by at most one edge. 

$iii)$ At most three edges can meet at any given vertex. 

\noindent
We will denote the vertices of $\Gamma$ by $P\in V(\Gamma)$ and the edges by $C\in E(\Gamma)$. 
To any such graph we attach a formal series of the form 
\eqn\cpB{
F_\Gamma(g_s,q) = \sum_{(\Upsilon,\phi)} 
{1\over |{\rm \Aut}(\Upsilon,\phi)|\prod_{e\in E(\Upsilon)}d_e}
C(\Upsilon,\phi) g_s^{2g(\Upsilon)-2}q^{\beta(\Upsilon,\phi)}}
with coefficients $C(\Upsilon,\phi)\in {\cal K}_T$
where we sum over (equivalence classes of) pairs $(\Upsilon, \phi)$ as above satisfying (1').
Here we define $\beta(\Upsilon,\phi)$ to be a formal linear combination of edges of $\Gamma$, 
$\beta(\Upsilon,\phi) = \sum_{e\in E(\Upsilon)} d_e\phi(e)$. 
$q = (q_1, \ldots , q_{|E(\Gamma)|})$ is a 
multisymbol associated to the edges of $\Gamma$, and $q^{\beta(\Upsilon,\phi)} 
= \prod_{e\in E(\Upsilon)} q_{\phi(e)}^{d_e}$. 

We decompose $\Gamma$ into subgraphs, 
by specifying a collection of points $Q_\alpha$, $\alpha=1,\ldots, M$ lying on distinct edges 
$C_1\ldots, C_M$ of $\Gamma$. No two points should lie on the same edge. Suppose we choose these
points so that $\Gamma$ is divided into several disconnected components $\Gamma_I$. 
The resulting graphs have more structure than the original graph $\Gamma$. A typical graph 
$\Gamma_I$ has two types of vertices:
old vertices inherited from $\Gamma$, and new univalent vertices resulting from the 
decomposition.  
We will also refer to old  and  new vertices as inner $V_i(\Gamma)$
and respectively outer $V_o(\Gamma)$ vertices. The edges of $\Gamma_I$ can 
also be classified in inner edges $E_i(\Gamma)$ -- which do not contain outer vertices -- and 
outer edges $E_o(\Gamma)$ -- which contain an outer vertex. 
Note that there is a unique outer edge $C_Q$ passing through each outer vertex $Q$. 
These graphs will be referred to as relative graphs. 

The decomposition of $\Gamma$ induces a similar decomposition of pairs $(\Upsilon, \phi)$. 
The points in the inverse image $\phi^{-1}(\{Q_\alpha\})$ divide $\Upsilon$ into 
disconnected graphs $\Upsilon_I$ which map to $\Gamma_I$ for each $I$. As before, a typical graph 
$\Upsilon_I$ has more structure than the original graph $\Upsilon$. The decomposition gives rise 
to a collection of new univalent vertices in addition to 
the ordinary vertices inherited from $\Upsilon$. Moreover, the edges and ordinary vertices of 
$\Upsilon_I$ inherit marking data from $\Upsilon$. The new vertices are unmarked. 
The new univalent vertices will be called outer vertices. The ordinary vertices 
will be referred to as inner vertices. An edge containing an outer vertex will be called outer 
edge. We denote by $V_{i,o}(\Upsilon_I)$, $E_{i,o}(\Upsilon_I)$ the set of inner/outer vertices 
and respectively edges. 
We also obtain a map of graphs $\phi_I: \Upsilon_I \ra \Gamma_I$ which maps the distinguished 
vertices of $\Upsilon_I$ to univalent vertices of $\Gamma_I$. 
To introduce some more terminology, 
we call the graphs $\Upsilon$ closed graphs while $\Upsilon_I$ will be 
called truncated graphs. 

Now, it is clear that all disconnected truncated graphs can be obtained by cutting closed 
graphs, and conversely, any closed graph can be obtained by gluing truncated graphs. 
We would like to use this idea in order to reconstruct the formal series \cpB\ from 
data associated to the graphs $\Gamma_I$. For each $\Gamma_I$ we need construct a formal series 
with coefficients in ${\cal K}_T$ by summing over equivalence classes of pairs 
$(\Upsilon_I, \phi_I)$. 
In order to write down such an expression we need to introduce some more notation. 
Given a pair $(\Upsilon_I,\phi_I)$ we define the genus  
\eqn\cpC{
g(\Upsilon_I)= 1- |V_i(\Upsilon_I)| + |E_i(\Upsilon_I)|+ 
\sum_{v\in V_i(\Upsilon_I)} g_v.}
and we denote by $h(\Upsilon_I)=|V_o(\Upsilon_I)|$ the number of outer vertices. 
For each univalent vertex of $\Gamma_I$, $Q\in V_o(\Gamma_I)$ we define a degree vector 
${\bf k}^I_Q(\Upsilon_I,\phi_I) 
=( k_{Q,1}^I, k_{Q,2}^I, \ldots )$ so that $k_{Q,m}^I$ is the number of outer edges 
of $\Upsilon_I$ projecting onto the outer ray $C_Q$ with degree $m$. 
${\bf k}_Q^I(\Upsilon_I, \phi_I)$ is an infinite 
vector with finitely many nonzero entries.  
Next, we have to introduce some formal variables keeping track of all this data. 
Let $q_I=(q_1^I,\ldots, q^I_{|E_i(\Gamma_I)|})$ 
and $\wq_I = (\wq_1^I, \ldots , \wq^I_{|E_o(\Gamma)|})$  associated to the inner and respectively 
outer edges of $\Gamma$. We define 
\eqn\cpD{
\beta(\Upsilon_I, \phi_I) = \sum_{e\in E_i(\Upsilon_I)} 
d_e^I \phi_I(e), \qquad {\wbeta}(\Upsilon_I,\phi_I) = \sum_{e\in E_o(\Upsilon_I)} 
d_e^J \phi_I(e)}
and 
\eqn\cpE{ 
q_I^{\beta(\Upsilon_I,\phi_I)} = \prod_{e\in E_i(\Upsilon_I)} {(q^I_{\phi_I(e)})}^{d^I_e}, \qquad 
{\wq}_I^{\wbeta(\Upsilon_I,\phi_I)} = \prod_{e\in E_o(\Upsilon_I)} {({\wq}^I_{\phi_I(e)})}^{d^I_e}.}
We also introduce formal variables $y_I=(y^I_{Q,m})_{m=1,\ldots,\infty, Q\in V_o(\Gamma_I)}$ 
and set 
\eqn\cpF{ 
y_I^{{\bf k}^I(\Upsilon_I, \phi_I)} =\prod_{Q\in V_i(\Gamma_I)} \prod_{m=1} 
{(y^I_{Q,m})}^{k^I_{Q,m}}.}
Then the formal series associated to $\Gamma_I$ takes the form 
\eqn\cpG{\eqalign{
& Z_{\Gamma_I}(g_s, q_I, \wq_I, y_I)\cr & = 
\sum_{(\Upsilon_I, \phi_I)} 
{C(\Upsilon_I, \phi_I)\over |\Aut(\Upsilon_I,\phi_I)|
\prod_{e\in E(\Upsilon_I)}d^I_e}g_s^{2g(\Upsilon_I)-2
+h(\Upsilon_I)} q_I^{\beta(\Upsilon_I,\phi_I)}{\wq}_I^{\wb(\Upsilon_I,\phi_I)}
y_I^{{\bf k}^I(\Upsilon_I, \phi_I)}\cr}}
where the coefficients $C(\Upsilon_I, \phi_I)\in {\cal K}_T$. Note that here we sum over 
all disconnected marked graphs $\Upsilon_I$, as opposed to \cpB\ where we sum over 
connected graphs. 

Now suppose we are given two relative graphs $\Gamma_I, \Gamma_J$. Choose a subset 
of outer vertices of $\Gamma_I$, $S_I \subset V_o(\Gamma_I)$, and a subset $S_J\subset 
V_o(\Gamma_J)$ so that $S_I\simeq S_J$. We glue $\Gamma_I$ and $\Gamma_J$ by choosing a 
bijection $\psi:S_I\ra S_J$, obtaining a relative graph $\Gamma_{IJ}$ with 
outer vertices $V_o(\Gamma_{IJ})=\left(V_o(\Gamma_I)\setminus S_I\right) \cup
\left(V_o(\Gamma_J)\setminus S_J\right)$ and outer edges 
$E_o(\Gamma_{IJ})=\left(E_o(\Gamma_I)\setminus E_I\right) \cup
\left(E_o(\Gamma_J)\setminus S_J\right)$ ($S_I, S_J$ can be equally well regarded as 
subsets of $E_o(\Gamma_I)$, $E_o(\Gamma_J)$.) 
The inner vertices of $\Gamma_{IJ}$ are
the union $V_i(\Gamma_{IJ})=V_i(\Gamma_I)\cup V_i(\Gamma_J)$. The inner edges of $\Gamma_{IJ}$ 
are given by $E_i(\Gamma_{IJ}) = E_i(\Gamma_{I})\cup E_i(\Gamma_J) \cup S$, where $S$ denotes 
the set of inner edges of $\Gamma_{IJ}$ obtained by gluing outer edges of $\Gamma_I, \Gamma_J$; 
$S\simeq S_I\simeq S_J$.  
To $\Gamma_{IJ}$ we associate a series 
\eqn\cpH{ \eqalign{
&Z_{\Gamma_{IJ}}(g_s, q_{IJ}, \wq_{IJ}, y_{IJ})\cr & = 
\sum_{(\Upsilon_{IJ}, \phi_{IJ})} {C(\Upsilon_{IJ}, \phi_{IJ})\over |\Aut(\Upsilon_{IJ},\phi_{IJ})|
\prod_{e\in E(\Upsilon_{IJ})}d^{IJ}_e}
g_s^{2g(\Upsilon_{IJ})-2
+h(\Upsilon_{IJ})}q_{IJ}^{\beta(\Upsilon_{IJ},\phi_{IJ})}{\wq}_{IJ}^{\wb(\Upsilon_{IJ},\phi_{IJ})}
y_{IJ}^{{\bf k}^{IJ}(\Upsilon_{IJ}, \phi_{IJ})}\cr}}
defined as above. The formal variables $q_{IJ}, \wq_{IJ}, y_{IJ}$ are defined in terms 
of $q_I, \wq_I, y_I$ and $q_J, \wq_J, y_J$ as follows

\eqn\cpHB{\vbox{\halign{ $#$ \hfill &\qquad  $#$ \hfill \cr
q^{IJ}_e = \left\{ \matrix{ q^I_e,\qquad & \hbox{if}\ e\in E_i(\Gamma_I) \cr & \cr 
                            q^J_e, \qquad & \hbox{if}\ e\in E_i(\Gamma_J) \cr & \cr
                            \wq^I_e\wq^J_e, \qquad & \hbox{if}\ e\in S\hfill}\right. &  
\wq^{IJ}_e =  \left\{ \matrix{ \wq^I_e, \qquad & \hbox{if}\ 
                               e\in E_o(\Gamma_I)\setminus S_I \cr & \cr 
                               \wq^J_e, \qquad & \hbox{if} \ 
                                e\in E_o(\Gamma_J)\setminus S_J \cr}\right.\cr
& y^{IJ}_Q = \left\{ \matrix{ y^I_Q, \qquad & \hbox{if}\ 
                              Q\in V_o(\Gamma_I)\setminus S_I \cr & \cr 
                            y^J_Q, \qquad & \hbox{if} \ 
                                Q\in V_o(\Gamma_J)\setminus S_J \hfill}\right..\cr}}}

\noindent We would like to represent \cpH\ as a pairing of the form 
\eqn\cpI{
Z_{\Gamma_{IJ}}(g_s, q_{IJ}, \wq_{IJ}, y_{IJ}) =\left\langle 
Z_{\Gamma_I}(g_s, q_I, \wq_I, y_I), Z_{\Gamma_J}(g_s, q_J, \wq_J, y_J)\right\rangle.}
based on gluing of pairs $(\Upsilon_I, \phi_I)$, $(\Upsilon_J, \phi_J)$.
Suppose $(Q, \psi(Q))\in S_I \times S_J$ are two outer vertices of $\Gamma_I, \Gamma_J$ 
identified in the gluing process. We denote by $C^I_Q, C^J_{\psi(Q)}$ the corresponding 
outer edges of $\Gamma_I, \Gamma_J$. A pair of truncated graphs $(\Upsilon_I, \phi_I)$ 
$(\Upsilon_J, \phi_J)$ can be glued if and only if the degrees of all outer edges 
of $\Upsilon_I$ projecting onto $C^I_Q$ match the degrees of all outer edges of 
$\Upsilon_J$ projecting onto $C^J_{\psi(Q)}$. Therefore 
two pairs $(\Upsilon_I, \phi_I)$ and $(\Upsilon_J, \phi_J)$ can be glued to form a
pair $(\Upsilon_{IJ}, \phi_{IJ})$ if and only if 
\eqn\cpIB{
{\bf k}^I_Q = {\bf k}^J_{\psi(Q)}\qquad  \forall\ Q\in S_I.}
Note that if this conditions is satisfied, one can identify any outer edge of $\Upsilon_I$ projecting 
to $C^I_Q$ 
to an outer edge of $\Upsilon_J$ projecting to $C^J_{\psi(Q)}$ as long as the degrees are equal.
This gives rise to 
(finitely) many different gluing combinations which may result in principle in 
different graphs $\Upsilon_{IJ}$.  
In fact, is not hard to work out the degeneracy of each pair $(\Upsilon_{IJ},\phi_{IJ})$ 
obtained by gluing 
a fixed pair $(\Upsilon_I, \phi_I)$,$(\Upsilon_J, \phi_J)$. By construction, we have a 
canonical embedding of groups 
\eqn\cpIC{
\Aut(\Upsilon_{IJ},\phi_{IJ}) \hookrightarrow \Aut(\Upsilon_I, \phi_I)\times 
\Aut(\Upsilon_J, \phi_J).} 
Given a particular gluing of the pairs $(\Upsilon_I, \phi_I)$,$(\Upsilon_J, \phi_J)$ one can 
obtain another gluing compatible with $(\Upsilon_{IJ}, \phi_{IJ})$ by separately acting with elements 
of $\Aut(\Upsilon_I, \phi_I),
\Aut(\Upsilon_J, \phi_J)$ on each pair. Apparently this gives rise to $|\Aut(\Upsilon_I, \phi_I)| 
|\Aut(\Upsilon_J, \phi_J)|$ gluing patterns resulting in the same pair $(\Upsilon_{IJ}, \phi_{IJ})$.
However, two of these patterns are equivalent if they are related by an element of 
$\Aut(\Upsilon_{IJ},\phi_{IJ})$ which acts simultaneously on 
the pairs $(\Upsilon_I, \phi_I)$,$(\Upsilon_J, \phi_J)$ through the embedding \cpIC.\ 
Therefore the degeneracy of $(\Upsilon_{IJ}, \phi_{IJ})$ is 
\eqn\cpID{
{|\Aut(\Upsilon_I, \phi_I)||\Aut(\Upsilon_J, \phi_J)|\over |\Aut(\Upsilon_{IJ},\phi_{IJ})|}}
Moreover, since the number of all possible gluing patterns of two fixed pairs 
$(\Upsilon_I, \phi_I)$,$(\Upsilon_J, \phi_J)$ is $\prod_{Q\in S_I} \prod_{m=1}^\infty (k^I_{Q,m})!$ 
we have the following formula 
\eqn\cpIE{
\prod_{Q\in S_I} \prod_{m=1}^\infty (k^I_{Q,m})! =\sum_{(\Upsilon_{IJ},\phi_{IJ})}
{|\Aut(\Upsilon_I, \phi_I)||\Aut(\Upsilon_J, \phi_J)|\over |\Aut(\Upsilon_{IJ},\phi_{IJ})|}}
where the sum is over all pairs $(\Upsilon_{IJ},\phi_{IJ})$ obtained by gluing 
$(\Upsilon_I, \phi_I)$,$(\Upsilon_J, \phi_J)$.
Since this argument is perhaps too abstract, some concrete examples may be clarifying at this point. 
It suffices to consider a very simple situation in which $\Gamma$ is a graph with two 
vertices, which is the case for example if $X$ is the total space of 
$\CO(-1)\oplus \CO(-1)\ra \IP^1$. 
We divide $\Gamma$ into two relative graphs by cutting the edge joining the two vertices as 
shown below. 
 
\ifig\gluingiii{Decomposition of the graph $\Gamma$ associated to $\CO(-1)\oplus\CO(-1)\ra \IP^1$.}
{\epsfxsize3.5in\epsfbox{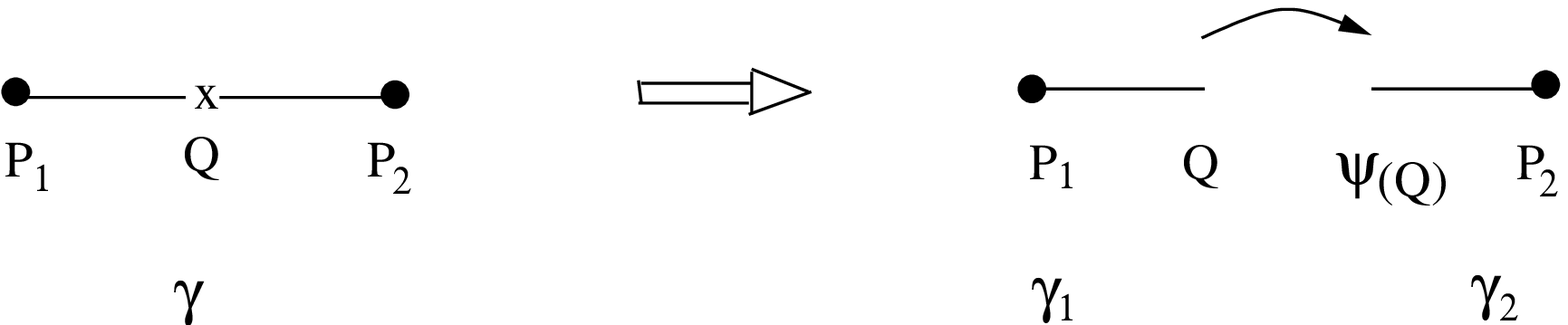}}

\noindent 
We consider two examples of gluing graphs represented in fig. 4 and fig. 5 below. 
In both cases, we draw the 
pair $(\Upsilon_I, \phi_I)$,$(\Upsilon_J, \phi_J)$ on the top row and all possible gluing patterns 
resulting in graphs $(\Upsilon_{IJ},\phi_{IJ})$ on the second row. 

\ifig\gluingi{First gluing example.}
{\epsfxsize4.8in\epsfbox{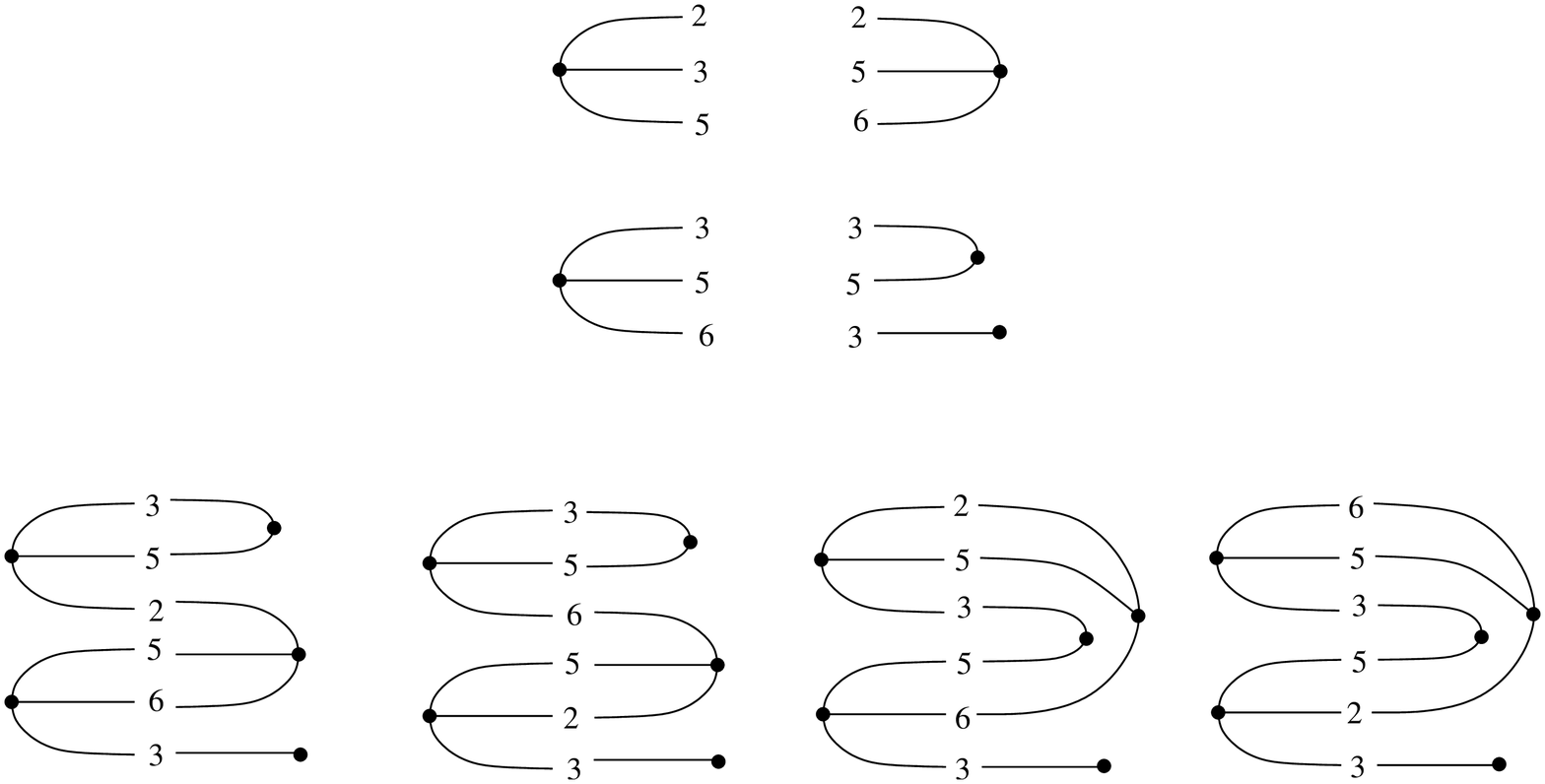}}

\noindent
In fig. 4 we have ${\bf k}^1_{Q}(\Upsilon_1,\phi_1)=(0,1,2,0,2,1,0,0,\ldots)$, hence 
$\prod_{m=1}^\infty (k^1_{Q,m})! = 2! \times 2! = 4$. $\Aut(\Upsilon_1,\phi_1)=\{1\}$, 
$\Aut(\Upsilon_2,\phi_2)=\{1\}$. There are four distinct gluing patterns, each resulting in
a connected closed string graph with trivial automorphism group. Therefore formula 
\cpIE\ holds. 
For the pair in fig. 5 we have $k^1_{Q,m}=(0,0,2,0,0,0,3,0,\ldots)$, hence 
$\prod_{m=1}^\infty (k^1_{Q,m})!= 2! \times 3! = 12$. $\Aut(\Upsilon_1,\phi_1)\simeq 
\Aut(\Upsilon_2,\phi_2) \simeq \CS_2\times \CS_2$. There are two distinct gluing patterns 
resulting in disconnected graphs with automorphism groups $\CS_2$ and respectively $\CS_2\times \CS_2$. 
Again the formula \cpIE\ holds. If the condition \cpIB\ is satisfied, one can easily show that 
\eqn\cpJ{\eqalign{ 
& 2g(\Upsilon_{IJ})-2+h(\Upsilon_{IJ}) = 2g(\Upsilon_I) -2 + h(\Upsilon_I) + 2g(\Upsilon_J)-2
+h(\Upsilon_J)\cr 
&\beta_{IJ}(\Upsilon_{IJ}, \phi_{IJ}) = \beta_I(\Upsilon_I,\phi_I) +\beta_J(\Upsilon_J,\phi_J) + 
\sum_{e\in S} d_e\phi_{IJ}(e)\cr
& \wbeta _{IJ}(\Upsilon_{IJ}, \phi_{IJ}) =\wbeta_I(\Upsilon_I,\phi_I) +\wbeta_J(\Upsilon_J,\phi_J) 
-\sum_{e\in S_I} d_e\phi_I(e) -\sum_{e\in S_J} d_e \phi_J(e)\cr
& {\bf k}^{IJ}_Q(\Upsilon_{IJ}, \phi_{IJ}) = \left\{\matrix{ 
{\bf k}^{I}_Q(\Upsilon_{I}, \phi_{I}),\qquad & \hbox{if}\ Q\in V_o(\Gamma_I)\setminus S_I\cr
{\bf k}^{J}_Q(\Upsilon_{J}, \phi_{J}),\qquad & \hbox{~if}\ Q\in V_o(\Gamma_J)\setminus S_J.\cr}
\right.\cr}}

\ifig\gluingii{Second gluing example.}
{\epsfxsize2.5in\epsfbox{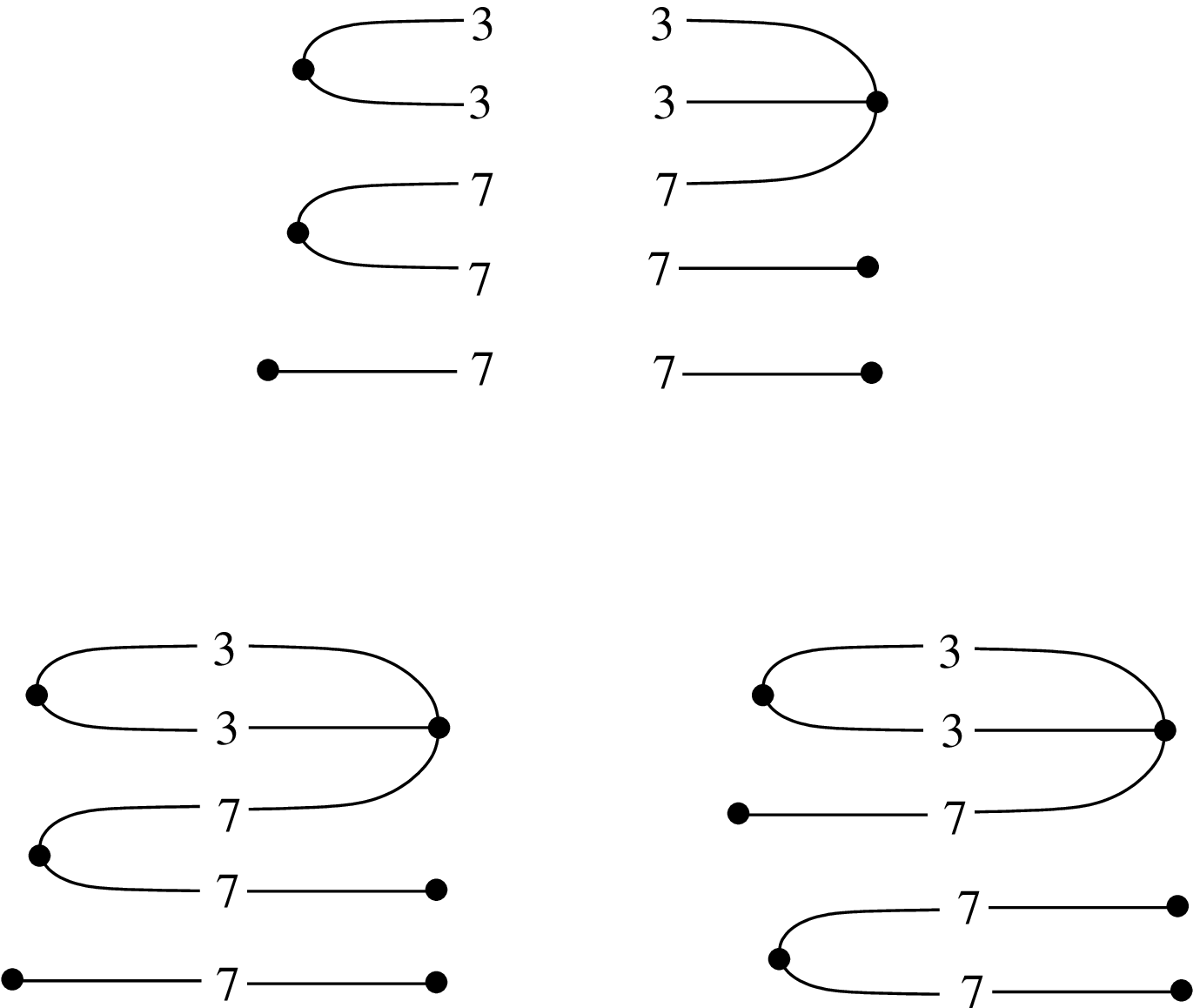}}

\noindent Using \cpHB\ and \cpI\ we find that the relation \cpJ\ imply
\eqn\cpKA{\eqalign{
& g_s^{2g(\Upsilon_{IJ})-2+h(\Upsilon_{IJ})} = g_s^{2g(\Upsilon_I) -2 + h(\Upsilon_I)}
g_s^{2g(\Upsilon_J)-2+h(\Upsilon_J)}\cr
& q_{IJ}^{\beta(\Upsilon_{IJ},\phi_{IJ})} \wq_{IJ}^{\wbeta(\Upsilon_{IJ},\phi_{IJ})} =
q_I^{\beta(\Upsilon_I,\phi_I)} \wq_I^{\beta(\Upsilon_I,\phi_I)} 
q_J^{\beta(\Upsilon_J,\phi_J)} \wq_J^{\beta(\Upsilon_J,\phi_J)}\cr}}
We define a formal pairing on $y$-variables by 
\eqn\cpK{\eqalign{
&\left\langle y_I^{{\bf k}^I(\Upsilon_I, \phi_I)}, 
y_J^{{\bf k}^J(\Upsilon_J, \phi_J)}\right\rangle = N({\bf k}^I_{S_I}(\Upsilon_I, \phi_I))
\prod_{Q\in  V_o(\Gamma_I)\setminus S_I} \prod_{m=1}^\infty {(y^I_{Q,m})}^{k^I_{Q,m}} 
\prod_{Q\in  V_o(\Gamma_J)\setminus S_J}\cr 
&\qquad\qquad\qquad\qquad\qquad\quad~\times\prod_{m=1}^\infty {(y^J_{Q,m})}^{k^J_{Q,m}}\prod_{Q\in S_I} \left(\prod_{m=1}^\infty 
m^{k^I_{Q,m}} (k^I_{Q,m})! \right)\delta( k^I_{Q,m}, k^J_{\psi(Q),m})\cr
& =  N({\bf k}^I_{S_I}(\Upsilon_I, \phi_I))y_{IJ}^{{\bf k}_{IJ}(\Upsilon_{IJ}, \phi_{IJ})} 
\prod_{Q\in S_I} 
\left(\prod_{m=1}^\infty m^{k^I_{Q,m}} (k^I_{Q,m})!\right) 
\delta({\bf k}^{I}_Q(\Upsilon_I,\phi_I), {\bf k}^J_{\psi(Q)}(\Upsilon_J,\phi_J)).\cr}}
where $N({\bf k}_{S_I}^I(\Upsilon_I, \phi_I))$ is a phase factor depending only on the winding vectors
of the outer edges which take part in the gluing process 
${\bf k}^I_{S_I}(\Upsilon_I,\phi_I) = \left({\bf k}^I_{Q}(\Upsilon_I,\phi_i)\right)_{Q\in S_I}$. 
The pairing 
is linear with respect to the other variables. 
The phase factor does not have a combinatoric explanation. It has 
to be included for geometric reasons explained in the next section. 
Using \cpKA\ and \cpK\ we can compute the right hand 
side of \cpI\
\eqn\cpM{\eqalign{ 
&\left\langle 
 Z_{\Gamma_I}(g_s, q_I, \wq_I, y_I), Z_{\Gamma_J}(g_s, q_J, \wq_J, y_J)\right\rangle \cr
&
=\sum_{(\Upsilon_I, \phi_I)} \sum_{(\Upsilon_J, \phi_J)}  
C(\Upsilon_I,\phi_I)C(\Upsilon_J,\phi_J)
g_s^{2g(\Upsilon_{IJ})-2+h(\Upsilon_{IJ})} 
q_{IJ}^{\beta(\Upsilon_{IJ},\phi_{IJ})} \wq_{IJ}^{\wbeta(\Upsilon_{IJ},\phi_{IJ})}
 y_{IJ}^{{\bf k}_{IJ}(\Upsilon_{IJ}, \phi_{IJ})} \cr
& \times{\prod_{Q\in S_I} 
\left(\prod_{m=1}^\infty m^{k^I_{Q,m}} (k^I_{Q,m})!\right)\over 
|\Aut(\Upsilon_I,\phi_I)|\left(\prod_{e\in E(\Upsilon_I)} d^I_e \right)|\Aut(\Upsilon_J,\phi_J)|
\left(\prod_{e\in E(\Upsilon_J)}d^J_e\right)}
\delta({\bf k}^{I}_Q(\Upsilon_I,\phi_I), {\bf k}^J_{\psi(Q)}(\Upsilon_J,\phi_J)).\cr}}

The $\delta$-symbol in the right hand side projects the sum onto pairs of graphs 
$(\Upsilon_I, \phi_I)$, $(\Upsilon_J, \phi_J)$ satisfying the gluing condition \cpIB.\ 
In order for the right hand side of \cpM\ to agree with \cpH,\ the coefficients 
$C(\Upsilon_I, \phi_I)$, $C(\Upsilon_J, \phi_J)$ must satisfy the gluing condition 
\eqn\cpN{
C(\Upsilon_I, \phi_I)C(\Upsilon_J, \phi_J)= N({\bf k}^I_{S_I}(\Upsilon_I,\phi_I))^*
C(\Upsilon_{IJ} , \phi_{IJ})}
for any pair $(\Upsilon_I, \phi_I)$, $(\Upsilon_J, \phi_J)$
satisfying \cpIB,\ and for any pair $(\Upsilon_{IJ},\phi_{IJ})$ obtained by 
gluing $(\Upsilon_I, \phi_I)$, $(\Upsilon_J, \phi_J)$.
Obviously, this condition is not of combinatoric nature. The coefficients in question 
must be specified by a particular geometric implementation of the gluing algorithm, 
which will be discussed in the next section. Here we will assume \cpN\ to be satisfied, 
and show that the pairing \cpM\ produces the expected result \cpH.\ 
Note that 
\eqn\cpNB{
{\prod_{Q\in S_I} \left(\prod_{m=1}^\infty m^{k^I_{Q,m}}\right)\over 
\left(\prod_{e\in E(\Upsilon_I)} d^I_e \right)
\left(\prod_{e\in E(\Upsilon_J)} d^J_e \right)} = 
{1\over \prod_{e\in E(\Upsilon_{IJ})} d^{IJ}_e}.}
This follows from the definition of ${\bf k}^I_Q(\Upsilon_I, \phi_I)$, using the 
gluing condition \cpIB.\ 

Using \cpIE,\ \cpN\ and \cpNB,\ in the right hand side of \cpM\ we find   
\eqn\cpP{
\eqalign{ 
&\left\langle 
 Z_{\Gamma_I}(g_s, q_I, \wq_I, y_I), Z_{\Gamma_J}(g_s, q_J, \wq_J, y_J)\right\rangle \cr
&=\sum_{(\Upsilon_{IJ}, \phi_{IJ})} 
{C(\Upsilon_{IJ}, \phi_{IJ})\over |\Aut(\Upsilon_{IJ},\phi_{IJ})|
\prod_{e\in E(\Upsilon_{IJ})}d^{IJ}_e}
g_s^{2g(\Upsilon_{IJ})-2
+h(\Upsilon_{IJ})} 
q_{IJ}^{\beta(\Upsilon_{IJ},\phi_{IJ})}{\wq}_{IJ}^{\wb(\Upsilon_{IJ},\phi_{IJ})}
y_{IJ}^{{\bf k}^{IJ}(\Upsilon_{IJ}, \phi_{IJ})}\cr}}
which is the expected result \cpH.\ This is our main gluing formula.

We would like to apply this gluing algorithm to the Gromov-Witten potential \cpB\
which is a sum over connected graphs $(\Upsilon, \phi)$. One can construct a generating 
functional for disconnected graphs by taking the exponential of \cpA.\
It is a standard fact that $Z_\Gamma(g_s,q)=\hbox{exp}(F_\Gamma(g_s,q))$ 
can be written as a sum over disconnected graphs 
\eqn\cpQ{
Z_\Gamma(g_s,q) = \sum_{(\Upsilon, \phi)} {1\over |{\rm \Aut}(\Upsilon,\phi)|\prod_{e\in E(\Upsilon)}d_e}
C(\Upsilon,\phi) g_s^{2g(\Upsilon)-2}q^{\beta(\Upsilon,\phi)}.}
One could use any decomposition of $\Gamma$ into relative graphs $\Gamma_I$. 
In particular we can cut $\Gamma$ along each edge, obtaining a collection of graphs $\Gamma_P$
labeled by vertices $P$ of $\Gamma$. Each $\Gamma_P$ has a inner vertex $P$ and three 
outer vertices. These graphs will be simply called vertices. 
The main 
problem is finding a natural geometric construction for the coefficients $C(\Upsilon_P, \phi_P)$ 
associated to $\Gamma_P$ satisfying the gluing conditions \cpN.\ 
This is the subject of the next section.

\newsec{Gluing Algorithm -- Geometry} 

This section consists of a geometric realization of the gluing algorithm.
We consider a decomposition of $\Gamma$ induced by intersecting the invariant 
curves $C_{rs}$ with (noncompact) lagrangian cycles $L_{rs}$ along circles $S_{rs}$;
$S_{rs}$ divides $C_{rs}$ into two discs with common boundary. To each circle $S_{rs}$ 
we can associate a point $Q_{rs}$ on the corresponding edge of $\Gamma$. The points 
$Q_{rs}$ divide 
$\Gamma$ into vertices as discussed in the last paragraph of the previous 
section. Each vertex represents a collection of (at most) three discs $D_i$, $i=1,2,3$
in $\IC^3$ with common origin. The boundaries of the discs are contained  
in  three lagrangian cycles $L_i$, 
$i=1,2,3$. Some vertices correspond to configurations of two or one 
discs, depending of the geometry. Those are special cases of the trivalent vertex.  

The main problem is finding a geometric construction for the generating functional \cpG\ 
so that the coefficients $C(\Upsilon_I, \phi_I)$ satisfy the gluing condition \cpN.\ 
A natural solution to this problem is suggested by string theory. One can wrap topological 
D-branes on the above lagrangian cycles, obtaining an open-closed topological string theory. 
Using the properties of this theory, one should be able to glue open string amplitudes 
obtaining closed string amplitudes. Therefore our generating functional should be 
the open string free energy associated to a collection of three lagrangian cycles 
in $\IC^3$ as above. 
The main problem at this point is that there is no complete mathematical formalism 
for open string Gromov-Witten invariants. 
We can approach the problem from the point of view of large $N$ duality and Chern-Simons 
theory as in \refs{\LMV,\OV}, or from an enumerative point of view as in \refs{\GZ,\KL,\LS,\ML,\Mii}. 
The first approach has been implemented in \AKMV,\ resulting in 
a gluing algorithm based on topological vertices. A topological vertex is the open 
string partition function of three lagrangian cycles in $\IC^3$ as predicted by 
large $N$ duality. Vertices can be naturally glued using a pairing very similar to \cpK.\ 

In this section we will take the second approach, constructing an open string generating functional
based on heuristic localization computations as in \refs{\GZ,\KL,\LS,\Mii}. The resulting expression 
can be written as a sum over open string graphs, as explained below, therefore it is tailor made 
for our construction. We will compare it in detail to the topological vertex in the next section. 

Let us start with some basic facts. 
The open string Gromov-Witten invariants count virtual 
numbers of maps 
$f:\Sigma \ra~\IC^3$, $f(\partial \Sigma) \subset L$ of fixed topological type, 
 where $\Sigma$ is a genus 
$g$ Riemann surface with $h$ boundary components. 
The $h$ boundary components 
are naturally divided into three groups, 
which are mapped to $L_1, L_2$ and respectively $L_3$. We will denote by 
$h_1, h_2, h_3$, $h_1+h_2+h_3=h$ the number of components in each group. 
We will also introduce three different sets of  indices 
$1\leq a_i \leq h_i$, $i=1,2,3$ in order 
to label the components in each group.
The topological type of the map $f$ is determined by three positive integers 
$(d_1,d_2, d_3)$ representing the degrees with respect to the three discs and 
three sets of winding numbers $n^i_{a_i}\geq 0$, $a_i=1,\ldots,h_i$, $i=1,2,3$. 
 In order to construct the generating functional, 
we introduce formal symbols $\wq_1,\wq_2,\wq_3$ keeping track of the degrees
and the formal variables $z_i=(z_{i,a_i})_{a_i=1,\ldots,\infty}$, 
$i=1,2,3$ keeping track of the winding numbers. 
\eqn\vpA{
F_\Lambda (g_s,q_i, z_i) = \sum_{g=0}^\infty \sum_{h=1}^\infty \sum_{d_i,n^i_{a_i}} 
g_s^{2g-2+h}C_{g,h_i}(d_i| n^i_{a_i}) \prod_{i=1}^3 \wq_i^{d_i}\prod_{a_i=1}^{h_i} 
z_{i,a_i}^{n^i_{a_i}}.}
Note that $C_{g,h_i}(d_i| n^i_{a_i})=0$ unless $d_i=\sum_{a_i=1}^{h_i} n^i_{a_i}$. 
This expression can be written in a more concise form if we introduce the winding vectors 
${\bf k}_{i} = (k_{i,m})_{m=1,\ldots ,\infty}$. Each vector has finitely many nonzero entry 
which count the number of boundary components of $\Sigma$ mapping to each lagrangian cycle 
with given 
winding number. More precisely $k^{i,m}$ represents the number of boundary components mapping 
to $L_i$ with winding number $m$. Note that we have 
$h_i=\sum_{m=1}^\infty k_{i,m}\equiv |{\bf k}_i|$, 
$d_i=\sum_{m=1}^\infty mk_{i,m}\equiv l({\bf k}_i)$. 
The coefficients $C_{g,h_i}(d_i|n^i_{a_i})$ are invariant under permutations 
of boundary components mapping to the same cycle $L_i$, hence they depend only on 
the ${\bf k}^i$. We can rewrite \vpA\ as 
\eqn\vpB{ 
F_\Lambda(g_s,q_i, y_i) = \sum_{g=0}^\infty g_s^{2g-2}\sum_{{\bf k}^i}C_g({\bf k}_i) \prod_{i=1}^3
g_s^{|{\bf k}_i|}q_i^{l({\bf k}_i)} \prod_{i=1}^3 y_i^{{\bf k}_i}}
where $y_i^{{\bf k}_i}=\prod_{m=1}^\infty y_{i,m}^{k_{i,m}}$. We have replaced the 
formal variables $z_{i}$ by new formal variables $y_i=(y_{i,m})_{m=1,\ldots,\infty}$ 
which keep track of the winding vectors ${\bf k}_i$. 

So far open string Gromov-Witten invariants have been rigorously constructed for a 
single disc in $\IC^3$ \refs{\ML} equipped with a torus action. There is an alternative 
computational 
definition \KL\ based on a heuristic application of the localization theorem of \GP\ to open 
string maps. Although not entirely rigorous, the second approach has been tested in many
physical situations with very good results \refs{\DFGi,\DFGii,\DF,\GZ,\KL,\LS,\Mii}. 
We will apply the same technique in order to construct the generating functional \vpB.\ 

Given a circle action $T\times \IC^3\ra ~\IC^3$ preserving 
$L$, one can compute $C_{g,h_i}(d_i| n^i_{a_i}) $ by localization. 
The fixed open string maps can be labeled 
by graphs \GZ\ by analogy with the closed string analysis of the previous section. 
The domain of a typical open string map is a union  
$\Sigma_{g,h} = \Sigma^0_g\cup \cup_{i=1}^3 \cup_{a_i=1}^{h_i}\Delta^i_{a_i}$ where $\Sigma_0$ 
is a closed prestable curve and $\Delta^i_{a_i}$ discs attached to $\Sigma^0_g$ at the 
marked points $p^i_{a_i}$. The data $(\Sigma^0_g, p^i_{a_i})$ must form a stable marked curve. 
The map $f:\Sigma\ra~\IC^3$ collapses $\Sigma^0_g$ to the origin $P=\{x_1=x_2=x_3=0\}$ 
and maps each $\Delta^i_{a_i}$ to $D_i$ with degree $n^i_{a_i}$. There are some special cases 
when $\Sigma^0_g$ is a point, which have to be treated separately (see appendix A.) 

Each fixed map is labeled by an open string graph with $h$ rays attached to a single vertex 
$v$. The vertex represents $\Sigma_0$, hence it is marked by the arithmetic genus $g_v$. 
The rays represent the discs $\Delta^{i}_{a_i}$, therefore they are marked by 
pairs $(i,n^i_{a_i})$. We will denote such marked graphs by $\Lambda$. The generating functional 
\vpB\ can be written as a sum over open string graphs 
\eqn\vpBB{
F_\Lambda (g_s,q_i,y_i) = \sum_{\Lambda} g_s^{2g(\Lambda)-2+h(\Lambda)} 
{1\over |\Aut(\Lambda)|\prod_{i=1}^3\prod_{a_i=1}^{h_i}n^i_{a_i}}C(\Lambda) \prod_{i=1}^3
q_i^{l_i(\Lambda)} \prod_{i=1}^3 y_i^{{\bf k}_i}}
where the notation is self-explanatory. For any graph $\Lambda$ we define $g(\Lambda), h(\Lambda), 
{\bf k}^i(\Lambda)$ to be the genus, number of rays and respectively $i$-th winding vector of the 
corresponding fixed map; $l_i(\Lambda) = l({\bf k}^i(\Lambda))$. 
The open string graphs are truncated graphs associated to the decomposition of 
$\Gamma$ in vertices, according to the terminology of the previous section.
The data of the map $\phi$ is encoded in the markings of the rays.  
The sum over graphs \vpBB\ is the local potential \cpG\ associated to a trivalent vertex. 
The coefficients $C(\Lambda)$, or, equivalently, $C_{g,h_i}(d_i|n^i_{a_i})$ are evaluated in 
appendix A. In the remaining part of this section we will show that they satisfy the 
gluing conditions \cpN.\ 

\subsec{Gluing Conditions} 

Let us consider a pair of trivalent vertices $\Gamma_r, \Gamma_s$ in the decomposition of 
$\Gamma$ which are glued to form a relative graph $\Gamma_{rs}$ as in fig. 6. 
The edge joining the two vertices corresponds to an invariant curve $C_{rs}$ on $X$. 
Let $(-a,-2+a)$, $a\in \IZ$ denote the type of $C_{rs}$. 

Consider two arbitrary open string graphs $\Lambda_r, \Lambda_s$ projecting to $\Gamma_r, \Gamma_s$
which satisfy the gluing condition ${\bf \kappa}^{r}_{1} = {\bf \kappa}^{s}_1$. Let 
$\Lambda_{rs}$ be a new open string graph projecting to $\Gamma_{rs}$ corresponding 
to an arbitrary gluing pattern of $\Lambda_r, \Lambda_s$. Here we want to prove the 
relation $C(\Lambda_{rs}) = N({\bf k}^r_{Q_{r1}}(\Lambda_r))C(\Lambda_r) C(\Lambda_s)$, 
where $N({\bf k}^r_{Q_{r1}}(\Lambda_r))$ is a phase factor. This is an essential 
condition for the gluing algorithm. 

Open string graphs can be evaluated using localization by analogy with closed 
string graphs. The open string coefficients have the following form 
\eqn\vpC{\eqalign{ 
C(\Lambda_I) = &\prod_{e_I\in E(\Lambda_I)}F_I(e_I)
\prod_{v_I\in V_i(\Lambda_I), (g_{v_I}, val(v_I)) =(0,2)}G_I(v_I)\cr 
& \times\prod_{v_I\in V_i(\Lambda_I), (g_{v_I}, val(v_I)) \neq (0,1), (0,2)} 
\int_{(\om_{g_{v_I},val(v_I)})_T} H_r(v_r)\cr}}
%C(\Lambda_s) = &\prod_{e_s\in E(\Lambda_s)}F_s(e_s)
%\prod_{v_s\in V_i(\Lambda_s), (g_{v_s}, val(v_s)) =(0,2)}G_s(v_s)\cr 
%& \times\prod_{v_s\in V_i(\Lambda_s), (g_{v_s}, val(v_s)) \neq (0,1), (0,2)} 
%\int_{(\om_{g_{v_s},val(v_s)})_T} H_s(v_s)\cr
%C(\Lambda_{rs}) = &\prod_{e_{rs}\in E(\Lambda_{rs})}F_{rs}(e_{rs})
%\prod_{v_{rs}\in V_i(\Lambda_{rs}), (g_{v_{rs}}, val(v_{rs})) =(0,2)}G_{rs}(v_{rs})\cr 
%& \times\prod_{v_{rs}\in V_i(\Lambda_{rs}), (g_{v_{rs}}, val(v_{rs})) \neq (0,1), (0,2)} 
%\int_{(\om_{g_{v_{rs}},val(v_{rs})})_T} H_{rs}(v_{rs})\cr}}
where the index $I$ takes values $I=r,s,rs$, and $F_{I}(e_{I})$, $G_{I}(v_{I})$, $H_{I}(v_{I})$
are edge and respectively vertex factors. The explicit expressions are computed in appendix A. 

\ifig\gluingiv{Gluing open string graphs.}
{\epsfxsize4.5in\epsfbox{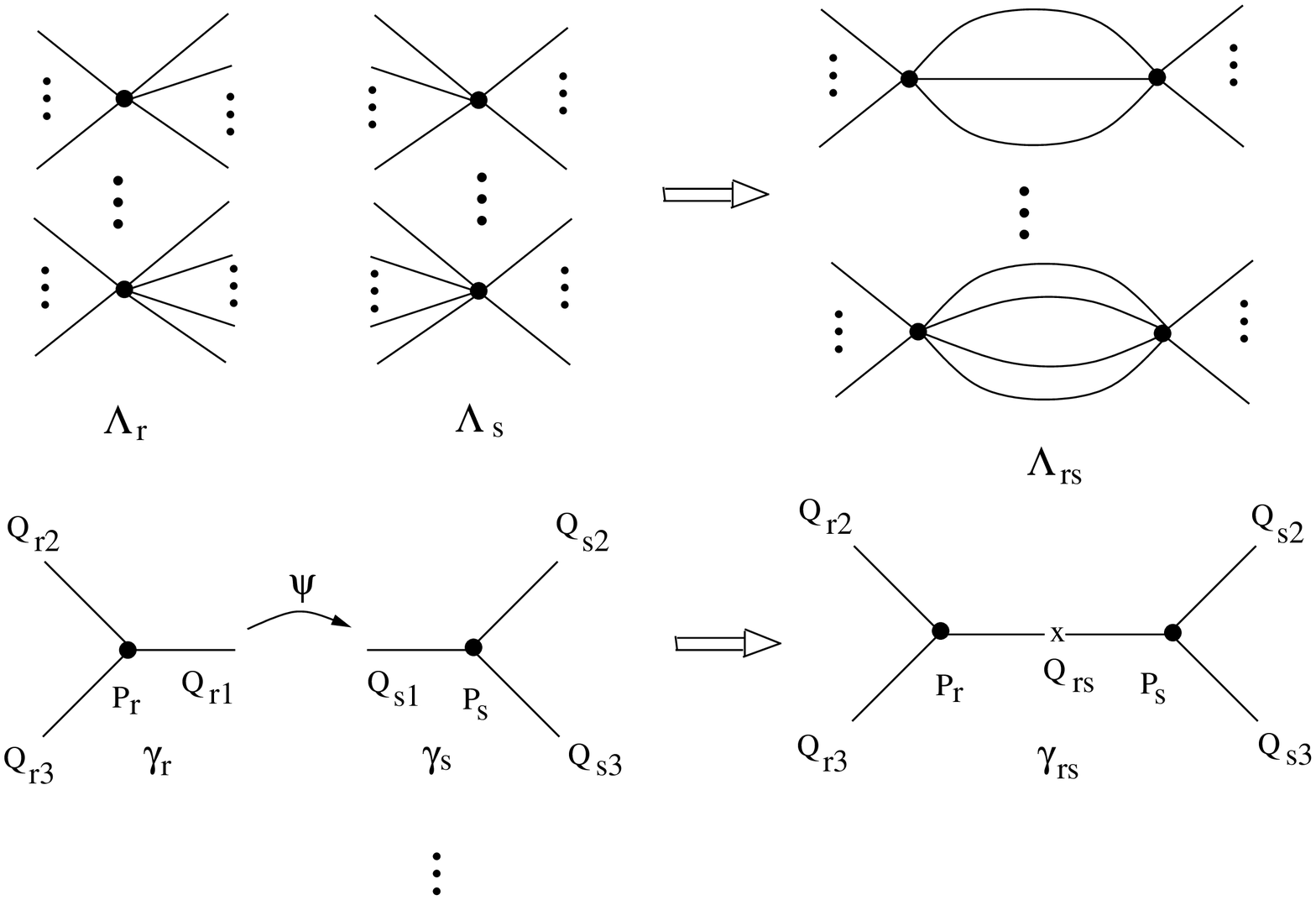}}

In order to simplify the notation, let us write $C(\Lambda_{r,s,rs})= C_e(\Lambda_{r,s,rs}) 
C_v(\Lambda_{r,s,rs})$ separating the edge and the vertex factors. 
Note that the set of inner vertices of $\Lambda_{rs}$ 
is $V_i(\Lambda_{rs})= V_i(\Lambda_r)\cup V_i(\Lambda_s)$. Moreover, the vertex factors 
$G_{r,s,rs}(v_{r,s,rs})$, $H_{r,s,rs}(v_{r,s,rs})$ are combinations 
of Hodge and Mumford classes determined 
by the marking data and the valence attached to a particular vertex. Since any inner vertex of 
$\Lambda_{rs}$ comes from an inner vertex in either $\Lambda_r$ or $\Lambda_s$, it follows that 
\eqn\vpD{\eqalign{
C_v(\Lambda_r) C_v(\lambda_s) = C_v(\Lambda_{rs}).}}
This leaves the edge factors. We have two types of edges. The outer edges associated to the 
univalent vertices $Q_{r2}, Q_{r3}$, $Q_{s2}, Q_{s3}$ are preserved by the gluing together with their 
markings. Therefore the corresponding edge factors remain trivially unchanged. The interesting edges 
are those associated to the univalent vertices $Q_{r1}, Q_{s1}$ which are identified in the 
gluing process. Geometrically, this corresponds to gluing two discs $D_{r1}, D_{s1}$ along their 
boundaries, obtaining the smooth rational curve $C_{rs}$. Before the gluing we have two products 
$C_{e1}(\Lambda_r)$, $C_{e1}(\Lambda_s)$ of open string edge factors. After the gluing we 
have a product $C_{e1}(\Lambda_{rs})$ of closed string edge factors. All three products have the 
same number of factors, one for each edge of $\Lambda_{rs}$ projecting to $C_{rs}$. 
Therefore the proof of the gluing conditions reduces to proving that 
\eqn\vpE{
F_{e_r}(\Lambda_r)F_{e_s}(\Lambda_s) = F_{e_{rs}}(\Lambda_{rs})}
for any pair of edges $e_r, e_s$ glued in the process. 
The edges $e_r, e_s$ correspond to two $T$-fixed open string maps $f_{r,s}:\Delta_{r,s} 
\ra D_{1r, 1s}$ with the same degree $d_r=d_s=d$. The edge $e_{rs}$ represents a $T$-fixed closed 
string map $f_{rs}: \Sigma_{rs}\simeq \IP^1\ra C_{rs}$ of degree $d$. Recall that we denote 
by $L_{rs}$ the lagrangian cycle which intersects $C_{rs}$ along the common boundary 
of $D_{1r}, D_{1s}$. 
A routine computation (see appendix B) shows that 
\eqn\vpG{
 F_{rs}(e_{rs}) = (-1)^{1+d(a-2)}  F_r(e_r)F_s(e_s).} 
Therefore we conclude that 
\eqn\vpH{
C(\Lambda_r) C(\Lambda_s)= \prod_{m=1}^\infty (-1)^{1+m(a-2)k^r_{Q_{r1},m}}C(\Lambda_{rs})}
This is the required gluing condition \cpN,\ in which the phase factor is a sign depending only on the 
winding vector ${\bf k}^r_{Q_{r1}}(\Lambda_r)={\bf k}^s_{Q_{s1}}(\Lambda_s)$. 

\newsec{Topological Vertex: Localization versus Chern-Simons}

In this section we compare the open string free energy \vpA\ with the topological vertex of 
\AKMV.\ 
The topological vertex is a generating functional for topological open string amplitudes derived 
from large $N$ duality. Each lagrangian cycle $L_i$ carries a flat unitary gauge field $A_i$ We denote by 
$V_i$ its holonomy around the boundary of the disc $D_i$. 
Then the topological vertex is given by the following expression \AKMV
\eqn\akmvvert{
Z=\sum_{{\bf k}^1,{\bf k}^2,{\bf k}^3}C_{{\bf k}^1{\bf k}^2{\bf k}^3}^{n_1n_2n_3}
\prod_{i=1}^3{1\over z_{{\bf k}^i}}
\Tr_{{\bf k}^i}V_i,
} 
where ${\bf k}^i$ are winding vectors, $z_{\bf k}=\prod_jk_j!j^{k_j}$ 
and $\Tr_{\bf k}V=\prod_{j=1}^{\infty}(\Tr V^j)^{k_j}$ 
and $n_1,n_2,n_3$ are the 
framing of the three legs of the vertex. The free energy derived from \akmvvert\ is to be 
compared with the results from 
localization (see Appendix A.) For the computation of the necessary Hodge integrals we have used 
Faber's Maple code \F . Below we 
list the coefficients of several terms with $h=3,g\leq 2$ in the expansion of the free energy.

$i)$$(\Tr V_1)^3$

\noindent Vertex result:
\eqn\vertone{\eqalign{
&{ig\over 6} n_1^2(n_1+1)^2-{ig^3\over 144}n_1^2(n_1+1)^2(8n_1^2+8n_1-9)
+{ig^5\over 11520}n_1^2(n_1+1)^2\big[8n_1(n_1+1)(13n_1^2\cr
&+13n_1-34)+189\big].}}
\noindent Localization result:
\eqn\locone{\eqalign{
&{ig}{\r_2^2\r_3^2\over 6\r_1^4}-{ig^3}{\r_2^2\r_3^2\over 144\r_1^6}\big[9\r_1(\r_2+\r_3)-8\r_2\r_3\big]
+{ig^5}{\r_2^2\r_3^2\over 11520\r_1^8}\big[-26\r_1^3(\r_2+\r_3)+163\r_1^2(\r_2+\r_3)^2\cr
&-272\r_1\r_2\r_3(\r_2+\r_3)+104\r_2^2\r_3^2\big].}}

$ii)$$(\Tr V_1)^2\Tr V_2$

\noindent Vertex result:
\eqn\vertoneprime{\eqalign{
&{ig\over 2}(n_1+1)^2-{ig^3\over 48}(n_1+1)^2(2n_1^2+4n_1+1)+{ig^5\over 11520}(n_1+1)^2\big[4n_1(n_1+2)(2n_1+1)\cr
&\times(2n_1+3)+3\big].}}

\noindent Localization result:
\eqn\loconeprime{\eqalign{
&ig{\r_3^2\over 2\r_1^2}-ig^3{\r_3^2\over 48\r_1^4\r_2}\big[\r_1^3+\r_1^2(4\r_2+\r_3)+2\r_1(2\r_2^2+\r_2\r_3)-2\r_2^2\r_3\big]
+{ig^5\over 11520}\cdot{\r_3^2\over \r_1^6\r_2^2}\big[\r_1^6+2\r_1^5\cr
&\times(19\r_2+5\r_3)+\r_1^4(112\r_2^2+58\r_2\r_3+5\r_3^4)+4\r_1^3\r_2(41\r_2^2+24\r_2\r_3+5\r_3^2)+4\r_1^2\r_2^2(23\r_2\cr
&+2\r_3)+8\r_1\r_2^3\r_3(11\r_2+7\r_3)+16\r_2^4\r_3^2\big].
}}

$iii)$$(\Tr V_1)^2\Tr V_2^2$

\noindent Vertex result:
\eqn\vertonetertz{\eqalign{
&{ig\over 2}\big[n_1^2n_2+n_1(2n_2-1)+2n_2-1)\big]-{ig^3\over 24}\big[n_1^4n_2+2n_1^3(2n_2-1)+n_1^2(2n_2^3+11n_2-6)\cr
&+2n_1(2n_2^3-3n_2^2+7n_2-3)+4n_2^3-6n_2^2+6n_2-2\big]+{ig^5\over 1440}\big[2n_1^6n_2+6n_1^5(2n_2-1)+5n_1^4\cr
&\times(2n_2^3+11n_2-6)+20n_1^3(2n_2^3-3n_2^2+7n_2-3)+n_1^2(6n_2^5+110n_2^3-180n_2^2+183n_2\cr
}}

$$-60)+2n_1(6n_2^5-15n_2^4+70n_2^3-90n_2^2+59n_2-15)+6(2n_2^5-5n_2^4+10n_2^3-10n_2^2+5n_2\hfill
-1)\big].$$

\noindent Localization result:
\eqn\loconetertz{\eqalign{
&{ig}{\r_3^2(2\r_3+\r_2)\over 2\r_1^2\r_2}-{ig^3}{\r_3^2(2\r_3+\r_2)\over 24\r_1^4\r_2^3}\big[2\r_1^3(\r_2-\r_3)
+2\r_1^2\r_2(2\r_2-\r_3)+2\r_1\r_2^3-\r_2^3\r_3\big]+{ig^5}\cr
&\times{\r_3^2(2\r_3+\r_2)\over 2880\r_1^6\r_2^5}\big[4\r_1^6(2\r_2-3\r_3)(3\r_2-\r_3)+4\r_1^5\r_2(23\r_2^2-32\r_2\r_3+\r_3^2)
+4\r_1^4\r_2^2(33\r_2^2\cr
&-32\r_2\r_3-6\r_3^2)+\r_1^3\r_2^3(87\r_2^2-73\r_2\r_3+16\r_3^2)+\r_1^2\r_2^4(23\r_2-19\r_3)(\r_2-\r_3)-2\r_1\r_2^5\r_3\cr
&\times(11\r_2+3\r_3)+4\r_2^6\r_3^2\big].
}}

$iv)$$\Tr V_1\Tr V_2\Tr V_3$

\noindent Vertex result:
\eqn\verttwo{
ig-{ig^3\over 24}+{ig^5\over 1920}.}
\noindent Localization result:
\eqn\loctwo{\eqalign{
&ig-{ig^3\over 24\r_1\r_2\r_3}\big[\r_1^2(\r_2+\r_3)+\r_1(\r_2^2+4\r_2\r_3+\r_3^2)+\r_2\r_3(\r_2+\r_3)\big]
+{ig^5\over 5760\r_1^2\r_2^2\r_3^2}\cr
&\times\big[5\r_1^4(\r_2+\r_3)^2+2\r_1^3(5\r_2^3+24\r_2^2\r_3+24\r_2\r_3^2+5\r_3^3)+\r_1^2(5\r_2^4+48\r_2^3
\r_3+102\r_2^2\r_3^2\cr
&+48\r_2\r_3^3+5\r_3^4)+2\r_1\r_2\r_3(5\r_2^3+24\r_2^2\r_3+24\r_2\r_3^2+5\r_3^3)+5\r_2^2\r_3^2(\r_2+\r_3)^2\big].}}

$v$)$\Tr V_1^2\Tr V_2\Tr V_3$

\noindent Vertex result:
\eqn\vertthree{\eqalign{
ig(2n_1+1)-{ig\over 6}(2n_1+1)(n_1^2+n_1+1)+{ig^5\over 120}(2n_1+1)\big[n_1(n_1+1)(n_1^2+n_1+3)+1\big].}}

\noindent Localization result:
\eqn\locthree{\eqalign{
&ig{\r_1+2\r_2\over \r_1}-ig^3{\r_1+2\r_2\over 24\r_1^3\r_2\r_3}\big[\r_1^3(\r_2+\r_3)+\r_1^2(\r_2^2+6\r_2\r_3+\r_3^2)
-4\r_2^2\r_3^2\big]
+ig^5{\r_1+2\r_2\over 5760\r_1^5\r_2^2\r_3^2}\cr
&\times\big[\r_1^6(\r_2+\r_3)^2+2\r_1^5(\r_2+\r_3)(5\r_2^2+29\r_2\r_3+5\r_3^2)
+\r_1^4(5\r_2^4+58\r_2^3\r_3+186\r_2^2\r_3^2+58\r_2\r_3^3\cr
&+5\r_3^4)-40\r_1^3\r_2^2\r_3^2(\r_2+\r_3)-8\r_1^2\r_2^2\r_3^2(9\r_2^2+46\r_2\r_3+9\r_3^2)-80\r_1\r_2^3\r_3^3(\r_2+\r_3)
+48\r_2^4\r_3^4\big].
}}

$vi$)$\Tr V_1^2\Tr V_2^2\Tr V_3$

\vfill\eject

\noindent Vertex result:
\eqn\vertfour{\eqalign{
&ig(n_1+2n_2+3n_1n_2)-{ig^3\over 24}\big[4n_1^3(3n_2+1)+24n_1^2n_2+3n_1(4n_2^3+4n_2^2+19n_2+1)+2n_2\cr
&\times(4n_2^2+15)\big]+{ig^5\over 1920}\big[16n_1^5(3n_2+1)+160n_1^4n_2+40n_1^3(4n_2^3+4n_2^2+19n_2+1)+80n_1^2n_2\cr
&\times(4n_2^2+15)+n_1(48n_2^5+80n_2^4+760n_2^3+120n_2^2+1167n_2+5)+32n_2^5+400n_2^3+410n_2)\big].
}}

\noindent Localization result:
\eqn\locfour{\eqalign{
&ig{(2\r_2+\r_1)(2\r_3+\r_2)\over \r_1\r_2}-ig^3{(2\r_2+\r_1)(2\r_3+\r_2)\over 24\r_1^3\r_2^3\r_3}\big[\r_1^3(\r_2^2-4\r_3^2)
+\r_1^2\r_2(\r_2^2+8\r_2\r_3-4\r_3^2)\cr
&-4\r_1\r_2^2\r_3^2-4\r_2^3\r_3^2\big]+ig^5{(2\r_2+\r_1)(2\r_3+\r_2)\over 24\r_1^5\r_2^5\r_3^2}\big[\r_1^6(5\r_2^4-72\r_2^2\r_3^2
-80\r_2\r_3^3+48\r_3^4)+\r_1^5\r_2\cr
&\times(10\r_2^4+78\r_2^3\r_3-80\r_2^2\r_3^2-576\r_2\r_3^3+16\r_3^4)+\r_1^4\r_2^2(5\r_2^4+78\r_2^3\r_3+240\r_2^2\r_3^2-400\r_2
\r_3^3\cr
&+168\r_3^4)-80\r_1^3\r_2^3\r_3^2(\r_2^2+5\r_2\r_3-4\r_3^2)-8\r_1^2\r_2^4\r_3^2(9\r_2^2+72\r_2\r_3-16\r_3^2)-16\r_1\r_2^5\r_3^3
(5\r_2\cr
&-r_3)+48\r_2^6\r_3^4\big].
}}

$vii$)$\Tr V_1^2\Tr V_2^2\Tr V_3^2$

\noindent Vertex result:
\eqn\vertfive{\eqalign{
&ig\big[2n_1(2n_2n_3+n_2+n_3)+2n_2n_3-1\big]-{ig^3\over 6}\big[2n_1^3(2n_2n_3+n_2+n_3)+3n_1^2(2n_2n_3-1)\cr
&+n_1(4n_2^3n_3+2n_2^3+6n_2^2n_3+4n_2n_3^3+6n_2n_3^2+40n_2n_3+13n_2+2n_3^3+13n_3)+2n_2^3n_3\cr
&-3n_2^2+2n_2n_3^3+13n_2n_3-3n_3^2-4\big]+{ig^5\over 360}\big[6n_1^5(2n_2n_3+n_2+n_3)+15n_1^4(2n_2n_3-1)\cr
&+10n_1^3
(4n_2^3n_3+2n_2^3+6n_2^2n_3+4n_2n_3^3+6n_2n_3^2+40n_2n_3+13n_2+2n_3^3+13n_3)+30n_1^2\cr
&\times(2n_2^3n_3-3n_2^2+2n_2n_3^3+13n_2n_3-3n_3^2-4)+n_1(12n_2^5n_3+6n_2^5+30n_2^4n_3+40n_2^3n_3^3\cr
&+60n_2^3n_3^2+400n_2^3n_3+130n_2^3+60n_2^2n_3^3+390n_2^2n_3+12n_2n_3^5+30n_2n_3^4+400n_2n_3^3\cr
&+390n_2n_3^2+1266n_2n_3+299n_2+6n_3^5+130n_3^3+299n_3)+6n_2^5n_3-15n_2^4+20n_2^3n_3^3\cr
&+130n_2^3n_3-90n_2^2n_3^2-120n_2^2+6n_2n_3^5+130n_2n_3^3+299n_2n_3-15n_3^4-120n_3^2-48\big].
}}

\noindent Localization result:
$$\eqalign{
&ig{(2\r_1+\r_3)(2\r_2+\r_1)(2\r_3+\r_2)\over \r_1\r_2\r_3}+ig^3{(2\r_1+\r_3)(2\r_2+\r_1)(2\r_3+\r_2)\over 6\r_1^3\r_2^3\r_3^3}
\big[\r_1^3(\r_2^3+\r_2^2\r_3\cr
&+\r_2\r_3^2+\r_3^3)+\r_1^2(\r_2^3\r_3-2\r_2^2\r_3^2+\r_2\r_3^3)+\r_1(\r_2^3\r_3^2+\r_2^2\r_3^3)+\r_2^3\r_3^3\big]+ig^5
{2\r_1+\r_3\over 360\r_1^5\r_2^5\r_3^5}\cr
&\times(2\r_2+\r_1)(2\r_3+\r_2)\big[\r_1^6(\r_2+\r_3)^2(3\r_2^4-5\r_2^3\r_3+15\r_2^2\r_3^2-5\r_2\r_3^3+3\r_3^4)+\r_1^5\r_2\r_3
}$$
\eqn\locfive{\eqalign{
&\times(\r_2+\r_3)(\r_2^4-41\r_2^3\r_3+36\r_2^2\r_3^2-41\r_2\r_3^3+\r_3^4)+\r_1^4\r_2^2\r_3^2
(8\r_2^4-5\r_2^3\r_3+54\r_2^2\r_3^2\cr
&-5\r_2\r_3^3+8\r_3^4)+5\r_1^3\r_2^3\r_3^3(\r_2+\r_3)(4\r_2^2-5\r_2\r_3+4\r_3^2)+8\r_1^2\r_2^4\r_3^4(\r_2^2-5\r_2\r_3+\r_3^2)\cr
&+\r_1\r_2^5\r_3^5(\r_2+\r_3)+3\r_2^6\r_3^6\big].
}}

\noindent 
Using the condition $\r_1+\r_2+\r_3=0$ derived in appendix A below, we find a complete agreement between 
the two expansions provided that the framing variables $n_i$ are related to the torus weights by 
\eqn\frams{
n_1={\r_2\over \r_1},~~n_2={\r_3\over\r_2},~~n_3={\r_1\over\r_3}.
}
It is easy to check that there is not choice of the torus weights rendering all $n_i$ integral. 
This may seem puzzling at first since the framing variables are traditionally integral in Chern-Simons 
theory. A first deviation from this rule was noticed in \DF\ in the context of large $N$ duality. 
Given the large $N$ duality origin of the topological vertex, the present result is not surprising. 
As pointed out in \DF,\ in order to obtain a consistent coupling of Chern-Simons theory and 
open string instanton corrections, the framing should be thought of as a formal variable.
Then all Chern-Simons expressions must be formally expanded as power series of these variables. 
The same strategy has been applied in this section, with very good results. We have also checked several terms 
with $h=4,g\leq 2$ and found agreement between the Chern-Simons and localization computations. 
In the light of this numerical evidence, we conjecture that the two generating 
functionals must agree to all orders. This result has been proved in \refs{\LLZi,\LLZii,\OP} 
for a univalent vertex. The trivalent vertex is an open problem. 

\appendix{A}{Open String Localization} 

Here we compute the generating functional \vpA\ using open string localization. 
Let $(x_1,x_2,x_3)$ be coordinates on $\IC^3$, and let 
\eqn\toractB{
x_1\ra e^{-i\rho_1\phi} x_1,\qquad x_2\ra e^{-i\rho_2\phi}x_2,\qquad 
x_3\ra e^{-i\rho_3\phi}x_3}
be an $S^1$ action. 
The lagrangian cycles $L_i$ are defined by the following equations 
\eqn\lagcyclesA{\eqalign{
& L_1:\qquad |x_1|=1, \qquad x_2 = \bx_3\bx_1\cr
& L_2:\qquad |x_2|=1,\qquad x_3 = \bx_1\bx_2\cr
& L_3:\qquad |x_3|=1,\qquad x_1 = \bx_2\bx_3.\cr}}
The $S^1$ action \toractB\ preserves $L=L_1\cup L_2\cup L_3$ if the weights 
$(\rho_1,\rho_2,\rho_3)$ satisfy 
\eqn\toractC{
\rho_1+\rho_2+\rho_3=0.} 
The three $S^1$-invariant discs ending on $L$ are given by 
\eqn\discsA{\eqalign{
& D_1:\qquad 0\leq |x_1| \leq 1,\qquad x_2=x_3=0\cr
& D_2:\qquad 0\leq |x_2|\leq 1,\qquad x_3=x_1=0\cr
& D_3:\qquad 0\leq |x_3| \leq 1,\qquad x_1=x_2=0.\cr}}

Let us describe the structure of an $S^1$ invariant map $f:\Sigma_{g,h}
\ra X$ with lagrangian boundary conditions on $L$. 
The map $f:\Sigma_{g,h} \ra X$ is constrained by stability and $S^1$ 
invariance. We give a  complete classification 
of all maps satisfying these two conditions, proceeding on a case by case 
basis. 

By $S^1$ invariance, $f$ must map $\Sigma_{g,h}$ onto the union
of three discs $D_1\cup D_2\cup D_3$. 
In the generic case, the domain must be a nodal bordered Riemann surface, 
consisting of a closed surface $\Sigma^0_g$ and three sets of discs 
$\Delta^i_{a_i}$, $i=1,2,3$ which are mapped to 
$D_1, D_2$ and respectively $D_3$. For future reference we will denote 
by $t_{ia_i}$ a coordinate on $\Delta^i_{a_i}$ centered at the origin. 
The discs are attached to $\Sigma^0_g$ by identifying the origins $t_{ia_i}=0$ 
to the marked points $p^i_{a_i}\in \Sigma^0_g$, so that 
we obtain a connected surface. The closed curve 
$\Sigma^0_g$ is mapped 
to the common origin $P$ of $D_1,D_2,D_3$. 
Stability further requires $(\Sigma^0_g, p^i_{a_i})$ 
to be a stable marked curve. We obtain several cases which should be spelled 
out in detail. 

$i)\ (g,h)=(0,1)$ In this case, the domain is a single disc, which can be 
mapped to $D_1, D_2$ or $D_3$. We have to distinguish accordingly three 
subcases 
\eqn\subcasesA{\eqalign{ 
& a)\ (g,h_i) =(0,1,0,0),\ (d_i| n^i_{a_i})=(d_1,0,0| d_1,0,0)\qquad
f:\Delta^1_1 \ra D_1,\ x_1 = t_{11}^{d_1} \cr
& b)\ (g,h_i) = (0,0,1,0),\ (d_i|n^i_{a_i})=(0, d_2,0|0,d_2,0)\qquad 
f:\Delta^2_1\ra D_2,\ x_2 = t_{21}^{d_2}\cr
& c)\ (g, h_i) =(0,0,0,1),\ (d_i| n^i_{a_i})=(0,0,d_3|0,0,d_3)\qquad 
f:\Delta^3_1\ra D_3, \ x_3=t_{31}^{d_3}.\cr}}
The automorphism group is $\Aut(f)\simeq \IZ/d_i$ where $i=1,2,3$. 

$ii)\ (g,h)=(0,2)$ The domain is a nodal (or pinched) annulus consisting 
of two discs with common origin. The two discs can be mapped either to 
the same disc $D_i$ in $X$ or to two different discs $D_i, D_j$, $i\neq j$. 
This yields again several subcases 
\eqn\subcasesB{\vbox{\halign{ $#$ \hfill &\qquad  $#$ \hfill \cr
a)\ (g,h_i) = (0,2,0,0),\ (d_i| n^i_{a_i})=(d_1,0,0| n^1_1, n^1_2,0,0)
 & f:\Delta^1_1 \cup \Delta^1_2 \ra D_1, \cr
& x_1 = t_{11}^{n^1_1} = t_{12}^{n^1_2}\cr
b)\ (g,h_i) = (0,0,2,0),\ (d_i| n^i_{a_i})=(0,d_2,0|0, n^2_1, n^2_2,0)
& f:\Delta^2_1 \cup \Delta^2_2 \ra D_2, \cr
& x_2 = t_{21}^{n^2_1} = t_{22}^{n^2_2}\cr
c)\ (g,h_i) = (0,0,0,2),\ (d_i| n^i_{a_i})=(d_1,0,0| 0,0,n^3_1, n^3_2)
& f:\Delta^3_1 \cup \Delta^3_2 \ra D_3, \cr
& x_3 = t_{31}^{n^3_1} = t_{32}^{n^3_2}\cr
d)\ (g,h_i)=(0,1,1,0),\ (d_i| n^i_{a_i})=(d_1,d_2,0|d_1,d_2,0) & 
f:\Delta^1_1\cup \Delta^2_1 \ra D_1\cup D_2,\cr
& x_1=t_{11}^{d_1}, x_2=t_{21}^{d_2}\cr
e)\ (g,h_i)=(0,1,0,1),\ (d_i| n^i_{a_i})=(d_1,0,d_3|d_1,0,d_3) & 
f:\Delta^1_1\cup \Delta^3_1 \ra D_1\cup D_3,\cr
& x_1=t_{11}^{d_1}, x_3=t_{31}^{d_3}\cr
f)\ (g,h_i)=(0,0,1,1),\ (d_i| n^i_{a_i})=(0,d_2,d_3|0,d_2,d_3) & 
f:\Delta^2_1\cup \Delta^3_1 \ra D_2\cup D_3,\cr
& x_2=t_{21}^{d_2}, x_3=t_{31}^{d_3}. \cr}}}
In the subcases $(a), (b)$ and $(c)$ the automorphism group is
\eqn\automB{
\Aut(f) = \left\{\matrix{ \IZ_{n^i_1}\times \IZ_{n^i_2},\qquad 
\quad \, i=1,2,3 \hfill &
\qquad \hbox{for}\ n^i_1\neq n^i_2\cr
\IZ_{n^i_1}\times \IZ_{n^i_2}\times \IZ/2, \ i=1,2,3 & \qquad 
\hbox{for}\  n^i_1=n^i_2\cr}\right.}
where the $\IZ/2$ factor in the second line is generated by a permutation 
of the two components of the domain. This is an automorphism if and only if 
$n^i_1= n^i_2$. 
For the remaining three cases, the automorphism group is 
\eqn\automC{
\Aut(f) = \IZ/d_i\times \IZ/d_{i+1}}
with the convention that $3+1$ is identified with $1$. Note that in this 
case, permuting the two components of the domain does not give rise 
to an automorphism even if $d_i=d_{i+1}$ for some $i=1,2,3$. 
To conclude the classification of all possible fixed loci, we have one more 
case which has been briefly mentioned earlier, namely the generic case 

$iii)\ (g,h)\neq (0,1), (0,2).$ The fixed map has the following form 
\eqn\subcasesC{
f:\Sigma^0_g\cup\left(\cup_{a_1=1}^{h_1} \Delta^1_{a_1}\right) \cup 
\left(\cup_{a_2=1}^{h_2} \Delta^2_{a_2}\right)\cup
\left(\cup_{a_3=1}^{h_3} \Delta^1_{a_3}\right)\ra D_1\cup D_2\cup D_3}
where $f(\Sigma^0_g) = P$ is a point, and 
\eqn\invmapA{\eqalign{
& x_1 = t_{11}^{n^1_1}=\ldots = t_{1h_1}^{n^1_{h_1}}\cr
& x_2 = t_{21}^{n^2_1}=\ldots = t_{2h_2}^{n^2_{h_2}}\cr
& x_2 = t_{31}^{n^3_1}=\ldots = t_{3h_3}^{n^3_{h_3}}.\cr}}
The marked Riemann surface
$(\Sigma_g^0, p^i_{a_i})$ must be a stable Deligne-Mumford 
curve.
In this case the automorphism group is a product between 
\eqn\automCA{
G=\prod_{i=1}^3\prod_{a_i=1}^{h_i} \IZ/n^i_{a_i}}
and a  subgroup 
\eqn\automD{\CP_1\times \CP_2\times \CP_2 \subset \CS_{h_1}
\times \CS_{h_2}\times \CS_{h_3}}
where $\CP_i$ permutes the marked points $\{p^i_{a_i}\}$ preserving the 
winding numbers $n^i_{a_i}$, $i=1,2,3$. In terms of the winding vectors ${\bf k}_i$, 
we have 
\eqn\automDB{
\CP_i \simeq \prod_{m=1}^\infty \CS_{k_{i,m}}.} 
Note that in all cases, the maps are fixed under the $T$ action provided that 
$T$ acts on the disc $\Delta^i_{a_i}$ as follows 
\eqn\toractC{
t_{ia_i}\ra e^{-i\phi\rho_i/n^i_{a_i}}t_{ia_i}.} 

The coefficients $C_{g,h_1,h_2,h_3}(d_i|n^i_{a_i})$ are computed by 
evaluating the contributions of the fixed points $(i)-(iii)$ to the 
virtual fundamental class. As usual with open string localization 
computations, the result is a homogeneous rational function of $(\rho_1, 
\rho_2, \rho_3)$ of degree zero. 
In the cases $(i)-(ii)$ above, the fixed locus in question is a point, 
therefore we have 
\eqn\locA{
C_{g,h_1,h_2,h_3}(d_i|n^i_{a_i}) ={1\over |\Aut(f)|} 
\int_{{pt}_{S^1}} {e({\IT}^2)\over e({\IT}^1)}.}
Here $e({\IT}^{1,2})$ denote the equivariant Euler classes of the 
terms in the tangent obstruction complex restricted to the fixed locus, 
and the integral represents equivariant integration along the fibers 
the map ${pt}_{S^1} \ra BS^1$. For the third case, we have similarly 
\eqn\locB{
C_{g,h_1,h_2,h_3}(d_i|n^i_{a_i}) ={1\over |\Aut(f)|} 
\int_{[\om_{g,h}]_{S^1}} {e({\IT}^2)\over e({\IT}^1)}.}

Next, we evaluate the contributions of the fixed points listed above starting 
with the generic case. Let 
$f_\partial : \Sigma_{g,h} \ra L$ denote the restriction
of $f:\Sigma_{g,h}\ra X$ to the boundary of $\Sigma_{g,h}$. The pair 
$\left(f^*T_X, f_\partial^*T_{L}\right)$ forms a Riemann-Hilbert bundle on 
$\left(\Sigma_{g,h},\partial \Sigma_{g,h}\right)$ and we will denote by 
$\CT_X$ the associated sheaf of germs of holomorphic sections. For future 
reference, we will denote by $\CT^i_{Xa_i}$ the restriction of $\CT_X$ 
to the disc $\Delta^i_{a_i}$. For simplicity, we will also denote the domain 
of $f$ by $\Sigma$, dropping the indices $(g,h)$. The closed surface 
$\Sigma_g^0$ will be similarly denoted by $\Sigma^0$, and the restriction 
$f|_{\Sigma^0}\equiv f^0$. 

The tangent-obstruction complex reads 
\eqn\tangobsA{
0\ra \Aut(\Sigma) \ra H^0(\Sigma, \CT_X) \ra {\IT}^1 \ra \Def(\Sigma) 
\ra H^1(\Sigma, \CT_X) \ra {\IT}^2\ra 0.} 
We denote the terms in this complex by $B_1, \ldots, B_6$ and the moving 
parts under the $S^1$ action by $B_1^m, \ldots ,B_6^m$.
Then \locB\ becomes  
\eqn\locC{
C_{g,h_1,h_2,h_3}(d_i|n^i_{a_i}) ={1\over |\Aut(f)|}
\int_{[\om_{g,h}]_{S^1}} {e(B_1^m)e(B_5^m) \over e(B^m_2)e(B^m_4)}.}
We have 
\eqn\locCA{
{e(B_1^m)\over e(B^m_4)} = {1\over e(\Def(\Sigma)^m)}}
since $\Aut(\Sigma)$ is generated in this case by 
$t_{ia_i} \partial_{t_{ia_i}}$ which are fixed by the $S^1$ action. 
The moving part of $\Def(\Sigma)^m$ is generated by deformations of the 
nodes, that is 
\eqn\locCB{
\Def(\Sigma)^m \simeq \oplus_{i=1}^3 \oplus_{a_i=1}^{h_i} 
T_{p^i_{a_i}}\Sigma^0 \otimes T_0\Delta^i_{a_i}.}
This yields 
\eqn\locCC{
e(\Def(\Sigma)^m)= \prod_{i=1}^3 \prod_{a_i=1}^{h_i} 
\left({\rho_i\over n^i_{a_i}}H -\psi_{ia_i}\right)}
where $H$ is the generator of $H^*(BS^1)$ and 
$\psi_{ia_i}\in H^*(\om_{g,h})$ are 
the Mumford classes associated to the marked points $p^i_{a_i}$ for 
$a_i=1,\ldots, h_i$, $i=1,2,3$. 
The other Euler classes in \locB\ can be evaluated using a (partial) 
normalization exact sequence 
\eqn\normseqA{
0\ra \CT_X \ra f_0^*T_X \oplus \oplus_{i=1}^3 \oplus_{a_i=1}^{h_i} \CT^i_{Xa_i}
\ra \oplus_{i=1}^{3}\oplus_{a_i=1}^{h_i} (T_X)_P\ra 0.}
The associated long exact sequence reads 
\eqn\longseqA{\eqalign{ 
0 & \ra H^0(\Sigma, \CT_X) \ra 
H^0(\Sigma^0,f_0^*T_X) \oplus \oplus_{i=1}^3 \oplus_{a_i=1}^{h_i} 
H^0(\Delta^i_{a_i},\CT^i_{Xa_i})\ra 
\oplus_{i=1}^{3}\oplus_{a_i=1}^{h_i} (T_X)_P\cr
&\ra H^1(\Sigma, \CT_X) \ra H^1(\Sigma^0,f_0^*T_X) \oplus 
\oplus_{i=1}^3 \oplus_{a_i=1}^{h_i} 
H^1(\Delta^i_{a_i},\CT^i_{Xa_i})\ra 0.\cr}}
We denote the terms in the complex \longseqA\ by $F_1, \ldots, F_5$. 
Then we have 
\eqn\locD{
{e(B^m_5)\over e(B^m_2)} = {e(F^m_5)e(F^m_3)\over e(F^m_2)}.}
In principle we have all the elements needed for the evaluation 
of the r.h.s. of \locD\ except the cohomology groups 
$H^{0,1}(\Delta^i_{a_i},\CT^i_{Xa_i})$. These groups can be computed as in 
\refs{\KL,\LS} or \DFGii\ obtaining the following expressions  
\eqn\tangobsB{\eqalign{
& H^0(\Delta^i_{a_i},\CT^i_{Xa_i})\simeq 
\left(\rho_i\right)\oplus \left({n^i_{a_i}-1\over n^i_{a_i}}\rho_i\right)
\oplus \ldots \oplus \left({1\over n^i_{a_i}}\rho_i\right)\oplus(0)_\IR\cr
& H^1(\Delta^i_{a_i},\CT^i_{Xa_i})\simeq \left(\rho_{i+1} + {1\over n^i_{a_i}}
\rho_i \right) \oplus \left( \rho_{i+1} + {2\over n^i_{a_i}}\rho_i\right) 
\oplus\ldots \oplus 
\left( \rho_{i+1} + {n^i_{a_i} -1 \over n^i_{a_i}}\rho_i\right),\cr}}
where $\rho_{3+1}$ should be identified with $\rho_1$. 
Now we can finish our Euler class computation
\eqn\eulerclsB{\eqalign{
& e(F^m_2) = H^{d+3}\prod_{i=1}^3 (\rho_i)^{d_i+1} \prod_{a_i=1}^{h_i} 
{\left(n^i_{a_i} -1\right)!\over \left(n^i_{a_i}\right)^{n^i_{a_i}-1}}\cr}}
\eqn\eulerclsC{\eqalign{
& e(F^m_5) = H^{d-h}\prod_{i=1}^3 c_g\left(\IE^*(\rho_i H)\right)
\prod_{a_i=1}^{h_i} \prod_{l=1}^{n^i_{a_i}-1}
\left(\rho_{i+1}+{l\over n^i_{a_i}}\right)\cr}}
\eqn\eulerclsD{\eqalign{
& e(F_3^m)= H^{3h} \prod_{i=1}^3\rho_i^{h}.\cr}}
Collecting all intermediate results, we are left with 
\eqn\locF{\eqalign{ 
C_{g,h_1,h_2,h_3}(d_i|n^i_{a_i})& ={1\over |\Aut(f)|}
\int_{[\om_{g,h}]_{S^1}} {e(B_1^m)e(B_5^m) \over e(B^m_2)e(B^m_4)}\cr
& = {1\over |\CP|}
{(\rho_1\rho_2\rho_3)^{h-1}\over \rho_1^{d_1}\rho_2^{d_2}\rho_3^{d_3}}
\prod_{i=1}^3\prod_{a_i=1}^{h_i} {\prod_{l=1}^{n^i_{a_i}-1} 
\left(n^i_{a_i}\rho_{i+1}+l\rho_i\right)\over (n^i_{a_i}-1)!}\cr
&\times\int_{[\om_{g,h}]_{S^1}} {H^{2h-3}\prod_{i=1}^3 c_g(\IE^*(\rho_i H))\over 
\prod_{i=1}^3 \prod_{a_i=1}^{h_i} 
\left({\rho_i}H - n^i_{a_i}\psi_{ia_i}\right)}.\cr}}
This represents the contribution of a generic fixed locus with $(g,h)\neq 
(0,1), (0,2)$ to the virtual fundamental class. Next we evaluate the
contributions of the fixed loci for the special cases $(i)-(ii)$.

In cases $(ia)-(ic)$, the domain is a single disc
$\Delta^i_1$ and $f:\Delta^i_1\ra D_i$ is a Galois 
cover of degree $d_i$. Using the same conventions and 
notations as above, we have
\eqn\euleriA{
{e(B^m_5)\over e(B^m_2)} = {H^{-1}\over \rho_i^{d_i}}\prod_{l=1}^{d_i-1}
{\left(d_i\rho_{i+1}+l\rho_i\right)\over (d_i-1)!}}
\eqn\euleriB{
{e(B^m_1)\over e(B^m_4)} = H{\rho_i\over d_i}.}
The last equation follows from the fact that for a disc $\Def(\Delta^i_1)^m$ 
is trivial, while $\Aut(\Delta^i_1)^m$ is generated by $\partial_{t_{i1}}$, 
which has weight ${\rho_i\over d_i}$. 
Taking into account the automorphism group, we obtain 
\eqn\lociA{\eqalign{
& C_{0,1}(d_1,0,0|d_1,0,0) = {1\over d_1^2} {1\over \rho_1^{d_1-1}}
{\prod_{l=1}^{d_1-1}(d_1\rho_2+l\rho_1)\over (d_1-1)!}\cr
& C_{0,1}(0,d_2,0|0,d_2,0) = {1\over d_2^2} {1\over \rho_2^{d_2-1}}
{\prod_{l=1}^{d_2-1}(d_2\rho_3+l\rho_2)\over (d_2-1)!}\cr
& C_{0,1}(0,0,d_3|0,0,d_3) = {1\over d_3^2} {1\over \rho_3^{d_3-1}}
{\prod_{l=1}^{d_3-1}(d_3\rho_1 +l\rho_3)\over (d_3-1)!}.\cr}}

Next we consider case $(ii)$. In the first three subcases $(iia)-(iic)$ 
the domain of $f$ is a nodal cylinder $\Sigma$ 
whose components are mapped in 
an invariant manner to the same disc in the target space $X$. It suffices to
do the computations for $(iia)$, since the remaining two cases are 
entirely analogous. We have to use a normalization sequence similar to 
\normseqA,\ except that the closed curve $\Sigma^0_g$ is absent. 
Therefore we have 
\eqn\normseqB{
0\ra \CT_X \ra \CT^{1}_{X1} \oplus \CT^{2}_{X1} \ra (T_X)_P \ra 0}
which yields the following long exact sequence 
\eqn\longseqB{\eqalign{
0 & \ra H^0(\Sigma, \CT_X) \ra H^0(\Delta^1_1, \CT^1_{X1}) \oplus 
H^0(\Delta^2_1, \CT^2_{X1}) \ra (T_X)_P\ra H^1(\Sigma, \CT_X)\cr
& \ra H^1(\Delta^1_1, \CT^1_{X1}) \oplus H^1(\Delta^2_1, \CT^2_{X1})
\ra 0\cr}}
whose terms will be denoted by $F_1, \ldots F_5$ as before. 
Then we can compute 
\eqn\lociiA{\eqalign{
{e(B^m_5)\over e(B^m_2)} & = {e(F^m_5)e(F^m_3)\over e(F^m_2)}\cr
& = \rho_1\rho_2\rho_3 H {\prod_{l=1}^{n^1_1-1}\left(n^1_1\rho_2+{l}\rho_1\right)
\prod_{l=1}^{n^1_2-1}\left(n^1_2\rho_2+{l}\rho_1\right)\over 
\rho_1^{d_1}(n^1_1-1)!(n^1_2-1)!}.\cr}}
The remaining factors are 
\eqn\lociiB{
{e(B^m_1)\over e(B^m_4)}= {1\over e(\Def(\Sigma)^m)} = 
{n^1_1n^1_2\over n^1_1+n^1_2}(\rho_1H)^{-1}}
since $\Aut(\Sigma)^m$ is trivial, and $\Def(\Sigma)^m$ is generated 
by deformations of the node
\eqn\lociiC{
\Def(\Sigma)^m \simeq T_0\Delta^1_1\otimes T_0\Delta^1_2.}
Substituting these expressions in \locA,\ we obtain the following result
\eqn\lociiD{\eqalign{
C_{0,2}(d_1,0,0|n^1_1,n^1_2,0,0) & = 
{1\over |\Aut(f)|} \int_{{pt}_{S^1}} 
{e(B^m_1)e(B^m_5)\over e(B^m_2)e(B^m_4)}\cr
& = {1\over |\CP|}{\rho_1\rho_2\rho_3\over \rho_1^{d_1+1}} 
{\prod_{l=1}^{n^1_1-1}\left(n^1_1\rho_2+{l}\rho_1\right)
\prod_{l=1}^{n^1_2-1}\left(n^1_2\rho_2+{l}\rho_1\right)\over 
(n^1_1-1)!(n^1_2-1)!(n^1_1+n^1_2)}.\cr}}
The results for $(iib)$ and $(iic)$ can be obtained by permuting 
the weights and the winding numbers 
\eqn\lociiE{\eqalign{
& C_{0,2}(0,d_2,0|0,n^2_1,n^2_2,0) ={1\over |\CP|} 
{\rho_1\rho_2\rho_3\over \rho_2^{d_2+1}} 
{\prod_{l=1}^{n^2_1-1}\left(n^2_1\rho_3+{l}\rho_2\right)
\prod_{l=1}^{n^2_2-1}\left(n^2_2\rho_3+{l}\rho_2\right)\over 
(n^2_1-1)!(n^2_2-1)!(n^2_1+n^2_2)}\cr
& C_{0,2}(0,0,d_3|0,0,n^3_1,n^3_2) = {1\over |\CP|}
{\rho_1\rho_2\rho_3\over \rho_3^{d_3+1}} 
{\prod_{l=1}^{n^3_1-1}\left(n^3_1\rho_1+{l}\rho_3\right)
\prod_{l=1}^{n^3_2-1}\left(n^3_2\rho_1+{l}\rho_3\right)\over 
(n^3_1-1)!(n^3_2-1)!(n^3_1+n^3_2)}.\cr}}
This leaves us with subcases $(iie)-(iif)$. Again, it suffices to do the 
computations only for $(iie)$. We have a map $f:\Delta^1_1\cup \Delta^2_1
\ra D_1\cup D_2$ which is a Galois cover of $D_1$ and respectively 
$D_2$ when restricted to
the components $\Delta^1_1,\Delta^2_1$. In this case the normalization exact 
sequence is 
\eqn\normseqC{ 
0\ra \CT_X \ra \CT^{1}_{X1} \oplus \CT^{2}_{X1} \ra (T_X)_P \ra 0.}
The associated long exact sequence reads 
\eqn\longseqC{\eqalign{
0&\ra H^0(\Sigma, \CT_X) \ra H^0(\Delta^1_1,\CT^{1}_{X1}) \oplus 
H^0(\Delta^2_1,\CT^{2}_{X1}) \ra (T_X)_P \ra H^1(\Sigma, \CT_X) \cr
& \ra  H^1(\Delta^1_1,\CT^{1}_{X1}) \oplus H^1(\Delta^2_1,\CT^{2}_{X1})
\ra 0.\cr}}
Repeating the previous steps, this yields 
\eqn\lociiF{\eqalign{
{e(B^m_5)\over e(B^m_2)} = 
H{\rho_1\rho_2\rho_3 \over \rho_1^{d_1}\rho_2^{d_2}}
{\prod_{l=1}^{d_1-1}\left(d_1\rho_2+l\rho_1\right)
\prod_{l=1}^{d_2-1}\left(d_2\rho_3+l\rho_2\right)\over (d_1-1)!(d_2-1)!}.}}
Moreover, the moving part of $\Def(\Sigma)$ is generated again by deformations 
of the node 
\eqn\lociiG{
\Def(\Sigma)^m \simeq T_0(\Delta^1_1)\otimes T_0(\Delta^2_1),}
which yields 
\eqn\lociiH{
{e(B^m_1)\over e(B^m_4)}={d_1d_2\over d_2\rho_1+d_1\rho_2}H^{-1}.}
Collecting all results we obtain the following expression 
\eqn\lociiI{
C_{0,2}(d_1,d_2,0|d_1,d_2,0)= 
{\rho_1\rho_2\rho_3 \over (d_2\rho_1+d_1\rho_2)\rho_1^{d_1}\rho_2^{d_2}}
{\prod_{l=1}^{d_1-1}\left(d_1\rho_2+l\rho_1\right)
\prod_{l=1}^{d_2-1}\left(d_2\rho_3+l\rho_2\right)\over (d_1-1)!(d_2-1)!}.}
For the remaining two cases, we can obtain the result by permuting the 
weights in \lociiI\
\eqn\lociiJ{\eqalign{
& C_{0,2}(0,d_2,d_3|0,d_2,d_3)= 
{\rho_1\rho_2\rho_3 \over (d_3\rho_2+d_2\rho_3)\rho_2^{d_2}\rho_3^{d_3}}
{\prod_{l=1}^{d_2-1}\left(d_2\rho_3+l\rho_2\right)
\prod_{l=1}^{d_3-1}\left(d_3\rho_1+l\rho_3\right)\over (d_2-1)!(d_3-1)!}\cr
& C_{0,2}(d_1,0,d_3|d_1,0,d_3)= 
{\rho_1\rho_2\rho_3 \over (d_1\rho_3+d_3\rho_1)\rho_3^{d_3}\rho_1^{d_1}}
{\prod_{l=1}^{d_3-1}\left(d_3\rho_1+l\rho_3\right)
\prod_{l=1}^{d_1-1}\left(d_1\rho_2+l\rho_1\right)\over (d_3-1)!(d_1-1)!}.\cr}}
To summarize this subsection, let us collect the results for $C_{g,h_i}(d_i|n^i_{a_i})$ 
\eqn\finalB{\eqalign{
& C_{0,1}(d_1,0,0|d_1,0,0) = {1\over d_1^2} {1\over \rho_1^{d_1-1}}
{\prod_{l=1}^{d_1-1}(d_1\rho_2+l\rho_1)\over (d_1-1)!}\cr
& C_{0,1}(0,d_2,0|0,d_2,0) = {1\over d_2^2} {1\over \rho_2^{d_2-1}}
{\prod_{l=1}^{d_2-1}(d_2\rho_3+l\rho_2)\over (d_2-1)!}\cr
& C_{0,1}(0,0,d_3|0,0,d_3) = {1\over d_3^2} {1\over \rho_3^{d_3-1}}
{\prod_{l=1}^{d_3-1}(d_3\rho_1 +l\rho_3)\over (d_3-1)!}\cr 
& C_{0,2}(d_1,0,0|n^1_1,n^1_2,0,0) = {1\over |\CP|}
{\rho_1\rho_2\rho_3\over \rho_1^{d_1+1}} 
{\prod_{l=1}^{n^1_1-1}\left(n^1_1\rho_2+{l}\rho_1\right)
\prod_{l=1}^{n^1_2-1}\left(n^1_2\rho_2+{l}\rho_1\right)\over 
(n^1_1-1)!(n^1_2-1)!(n^1_1+n^1_2)}\cr
& C_{0,2}(0,d_2,0|0,n^2_1,n^2_2,0) = {1\over |\CP|}
{\rho_1\rho_2\rho_3\over \rho_2^{d_2+1}} 
{\prod_{l=1}^{n^2_1-1}\left(n^2_1\rho_3+{l}\rho_2\right)
\prod_{l=1}^{n^2_2-1}\left(n^2_2\rho_3+{l}\rho_2\right)\over 
(n^2_1-1)!(n^2_2-1)!(n^2_1+n^2_2)}\cr
& C_{0,2}(0,0,d_3|0,0,n^3_1,n^3_2) = {1\over |\CP|}
{\rho_1\rho_2\rho_3\over \rho_3^{d_3+1}} 
{\prod_{l=1}^{n^3_1-1}\left(n^3_1\rho_1+{l}\rho_3\right)
\prod_{l=1}^{n^3_2-1}\left(n^3_2\rho_1+{l}\rho_3\right)\over 
(n^3_1-1)!(n^3_2-1)!(n^3_1+n^3_2)}\cr
& C_{0,2}(d_1,d_2,0|d_1,d_2,0)= 
{\rho_1\rho_2\rho_3 \over (d_2\rho_1+d_1\rho_2)\rho_1^{d_1}\rho_2^{d_2}}
{\prod_{l=1}^{d_1-1}\left(d_1\rho_2+l\rho_1\right)
\prod_{l=1}^{d_2-1}\left(d_2\rho_3+l\rho_2\right)\over (d_1-1)!(d_2-1)!}\cr
& C_{0,2}(0,d_2,d_3|0,d_2,d_3)= 
{\rho_1\rho_2\rho_3 \over (d_3\l_2+d_2\rho_3)\rho_2^{d_2}\rho_3^{d_3}}
{\prod_{l=1}^{d_2-1}\left(d_2\rho_3+l\rho_2\right)
\prod_{l=1}^{d_3-1}\left(d_3\rho_1+l\rho_3\right)\over (d_2-1)!(d_3-1)!}\cr
& C_{0,2}(d_1,0,d_3|d_1,0,d_3)= 
{\rho_1\rho_2\rho_3 \over (d_1\rho_3+d_3\rho_1)\rho_3^{d_3}\rho_1^{d_1}}
{\prod_{l=1}^{d_3-1}\left(d_3\rho_1+l\rho_3\right)
\prod_{l=1}^{d_1-1}\left(d_1\rho_2+l\rho_1\right)\over (d_3-1)!(d_1-1)!}\cr
& C_{g,h_1,h_2,h_3}(d_i|n^i_{a_i}) 
= {1\over |\CP|}{(\rho_1\rho_2\rho_3)^{h-1}\over \rho_1^{d_1}\rho_2^{d_2}\rho_3^{d_3}}
\prod_{i=1}^3\prod_{a_i=1}^{h_i} {\prod_{l=1}^{n^i_{a_i}-1} 
\left(n^i_{a_i}\rho_{i+1}+l\rho_i\right)\over (n^i_{a_i}-1)!}\cr
&\qquad\qquad \qquad \qquad\ \
\times\int_{[\om_{g,h}]_{S^1}} {H^{2h-3}\prod_{i=1}^3 c_g(\IE^*(\rho_i H))\over 
\prod_{i=1}^3 \prod_{a_i=1}^{h_i} 
\left({\rho_i}H - n^i_{a_i}\psi_{ia_i}\right)}.\cr}}

\appendix{B}{The Gluing Condition for Open String Graphs}

In this appendix we prove the gluing formula \vpE\ for an arbitrary invariant curve 
$C_{rs}$ of type $(-a, -2+a)$. Let $\CU$ be an open neighborhood of $C_{rs}$ in $X$ 
which can be covered by two smooth coordinate patches 
$\CU_r, \CU_s$ with coordinates $(x_1,x_2,x_3)$ and $(y_1,y_2,y_3)$. 
The local coordinates are chosen 
so that $(x_1, y_1)$ are affine coordinates on $\IP^1$, while $(x_2,x_3)$ 
and respectively $(y_2,y_3)$ are normal coordinates in the two patches. 
Therefore the transition functions are 
\eqn\transfctA{
y_1={1\over x_1},\qquad y_2 = x_1^ax_2,\qquad y_3=x_1^{-2+a}x_3.} 
In terms of local coordinates, the torus action reads 
\eqn\toractD{\vbox{\halign{ $#$ \hfill &\qquad  $#$ \hfill &\qquad  $#$ \hfill \cr
x_1\ra e^{-i\rho_1\phi}x_1, & x_2\ra e^{-i\rho_2\phi}x_2, & x_3\ra e^{-i\rho_3\phi}x_3, \cr
y_1\ra e^{i\rho_1\phi}y_1,  & y_2\ra e^{-i(a\rho_1+\rho_2)\phi}y_2, & y_3\ra e^{-i((-2+a)\rho_1+\rho_3)\phi}y_3.\cr}}}
Note that the local form of the action in the patch $\CU_2$ is determined
by the local action in $\CU_1$ and the transitions functions 
\transfctA.\ 
We denote by 
\eqn\fixedpts{
P_r:\quad x_1=x_2=x_3=0,\qquad P_s:\quad y_1=y_2=y_3=0}
the fixed points of the torus action. 

There are five lagrangian cycles in $\CU_{rs}=\CU_r\cup \CU_s$ given by 
\eqn\lagcyclesB{\eqalign{
& L_{r1}:\quad |x_1|=1, \quad x_2 = \bx_3\bx_1,\qquad 
L_{s1}:\quad |y_1|=1,\quad y_2=\by_3\by_1\cr
& L_{r2}:\quad |x_2|=1,\quad x_3 = \bx_1\bx_2,\qquad
L_{s2}:\quad |y_2|=1,\quad y_3=\by_1\by_2\cr
& L_{r3}:\quad |x_3|=1,\quad x_1 = \bx_2\bx_3,\qquad 
L_{s3}:\quad |y_3|=1,\quad y_1=\by_2\by_3.\cr}}
Note however that $L_{r1}=L_{s1}$ are identical cycles.
This can be seen using the transition functions \transfctA.\ 
We also have six discs ending on the lagrangian cycles. For the present purposes we will 
consider only two of them 
\eqn\discsB{\eqalign{
&D_{r1}:\qquad 0\leq |x_1| \leq 1,\qquad x_2=x_3=0\cr
&D_{s1}:\qquad 0\leq |y_1|\leq 1,\qquad y_2=y_3=0.\cr}}

We consider open string fixed maps 
\eqn\invmapsA{
f_r:\Delta_r\ra D_{r1},\qquad f_s:\Delta_s\ra D_{s1}}
of the same degree $n$ which yield a degree $n$ closed string map  
\eqn\invmapsB{ 
f_{rs}:\Sigma_{rs} \ra C_{rs}} 
by gluing. 
We denote by $\CT_{Xr}, \CT_{Xs}$ the corresponding
Riemann-Hilbert bundles on $\Delta_r, \Delta_s$ as in appendix A. 
The edge factors are
\eqn\edgefactorsA{\eqalign{
& F_r({e_r}) = {e(H^1(\Delta_r, \CT_{Xr})^m) e(\Aut(\Delta_r)^m)\over 
e(H^0(\Delta_r, \CT_{Xr})^m)}\cr
& F_s({e_s}) = {e(H^1(\Delta_s, \CT_{Xs})^m) e(\Aut(\Delta_s)^m)\over 
e(H^0(\Delta_s, \CT_{Xs})^m)}\cr
& F_{rs}(e_{rs}) = {e(H^1(\Sigma_{rs}, f_{rs}^*T_X)^m) e(\Aut(\Sigma_{rs})^m)\over 
e(H^0(\Sigma_{rs}, f^*T_X)^m)}.\cr}}
Let us compute the equivariant 
Euler classes in \edgefactorsA.\ For discs we can copy the results 
of the previous section (eqn.\tangobsB ) taking into account the 
local form of the $S^1$ action \toractD\ 
\eqn\edgefactorsB{\eqalign{
& {e(H^1(\Delta_r, \CT_{Xr})^m)\over 
e(H^0(\Delta_r, \CT_{Xr})^m)} = 
H^{-1}{\prod_{k=1}^{n-1}\left(n\rho_2+k\rho_1\right)\over \rho_1^n(n-1)!}\cr
 & {e(H^1(\Delta_s, \CT_{Xs})^m)\over 
e(H^0(\Delta_s, \CT_{Xs})^m)} = H^{-1}
{\prod_{k=1}^{n-1}\left(n(a\rho_1+\rho_2)-k\rho_1\right)\over (-\rho_1)^n(n-1)!}.\cr}}

In order to compute the edge factor for $f_{rs}:\Sigma_{rs}\ra X$ we have to use 
the short exact sequence of the image 
\eqn\shortseqA{
0\ra T_{C_{rs}} \ra {T_X}|_{C_{rs}} \ra N_{C_{rs}/X}\ra 0.} 
This induces a short exact sequence on the domain\foot{In order for the first 
and last term to make sense, we have to think of $f_{rs}$ 
as a map to $C_{rs}$ instead of $X$. This is a slight abuse of notation.}
\eqn\shortseqB{
0\ra f_{rs}^*T_{C_{rs}}\ra f_{rs}^*T_X \ra f_{rs}^*N_{C_{rs}/X}\ra 0.}
The associated long exact sequence reads 
\eqn\longseqD{\eqalign{
0 & \ra H^0(\Sigma_{rs}, f_{rs}^*T_{C_{rs}}) \ra H^0(\Sigma_{rs}, f_{rs}^*T_X) \ra 
H^0(\Sigma_{rs},f_{rs}^*N_{C_{rs}/X})\ra \cr
&\ra H^1(\Sigma_{rs}, f_{rs}^*T_{C_{rs}}) \ra H^1(\Sigma_{rs}, f_{rs}^*T_X) \ra 
H^1(\Sigma_{rs},f_{rs}^*N_{C_{rs}/X})\ra 0\cr}}
which shows that 
\eqn\edgefactorsC{ 
{e(H^1(\Sigma_{rs}, f_{rs}^*T_X)^m)\over 
e(H^0(\Sigma_{rs}, f_{rs}^*T_X)^m)} = {e(H^1(\Sigma_{rs}, f_{rs}^*T_{C_{rs}})^m)
e(H^1(\Sigma_{rs}, f_{rs}^*N_{C_{rs}/X})^m)\over 
e(H^0(\Sigma_{rs}, f_{rs}^*T_{C_{rs}})^m)e(H^0(\Sigma_{rs}, f_{rs}^*N_{C_{rs}/X})^m)}.}
Now let us compute the cohomology groups. Recall that $T_{C_{rs}} \simeq \CO(2)$ 
and $N_{C_{rs}/X} \simeq \CO(-a)\oplus\CO(a-2)$. Without loss of generality, 
we can assume $a\geq 1$. Then we have 
\eqn\edgefactorsD{\eqalign{
& H^0(\Sigma_{rs}, f_{rs}^*T_{C_{rs}}) \simeq H^0(\IP^1, \CO(2n))\cr
& H^0(\Sigma, f_{rs}^*N_{C_{rs}/X}) \simeq H^0(\IP^1, \CO(-an) \oplus \CO((a-2)n)).\cr}}
Moreover, using Kodaira-Serre duality 
\eqn\KSduality{\eqalign{
& H^1(\Sigma_{rs}, f_{rs}^*T_{C_{rs}}) \simeq H^0(\Sigma_{rs}, f_{rs}^*(T_{C_{rs}}^*)
\otimes \omega_{\Sigma_{rs}})^* 
\simeq H^0(\IP^1, \CO(-2-2n))^* = 0 \cr
& H^1(\Sigma_{rs}, f_{rs}^*N_{C_{rs}/X}) \simeq 
H^0(\Sigma_{rs}, f_{rs}^*(N_{C_{rs}/X}^*)\otimes \omega_{\Sigma_{rs}})^* \simeq 
H^0(\IP^1, \CO(an-2)\oplus \CO((2-a)n-2))^*.\cr}}
We can write down explicit generators as follows 
\eqn\cohomgenA{\eqalign{
& H^0(\Sigma_{rs}, f_{rs}^*T_{C_{rs}}):\qquad \partial_{x_1}, t\partial_{x_1},\ldots, 
t^{2n}\partial_{x_1}\cr
& H^0(\Sigma_{rs}, f^*N_{C_{rs}/X}): \qquad 
\left\{\matrix{ 0,\hfill & \hbox{if}\ a=1\cr 
\partial_{x_3}, t\partial_{x_3}, \ldots t^{(a-2)n}\partial_{x_3} & 
\hbox{if} \ a\geq 2\cr}\right.\cr
& H^1(\Sigma_{rs}, f^*N_{C_{rs}/X})^*:\qquad 
\left\{\matrix{ dx_2dt, tdx_2dt, \ldots, t^{n-2}dx_2dt, & \hbox{if}\ a=1\cr
dx_3dt, tdx_3dt, \ldots, t^{n-2}dx_3dt &  \cr
dx_2dt, tdx_2dt,\ldots, t^{an-2}dx_2dt \hfill & \hbox{if}\ a\geq 2.\cr}
\right.\cr}}
where $t$ is an affine coordinate of $\Sigma$ so that $f:\Sigma \ra C$ 
is locally given by $x_1=t^n$. 
In terms of representations of $S^1$, we have 
\eqn\cohomgenB{\eqalign{
& H^0(\Sigma_{rs}, f^*T_{C_{rs}})\simeq \oplus_{k=-n}^n \left({k\over n} \rho_1\right)\cr
& H^0(\Sigma_{rs}, f^*N_{C_{rs}/X})\simeq \left\{\matrix{0,\hfill& \hbox{if}\ 
a=1\cr 
\oplus_{k=0}^{n(a-2)} 
\left(\rho_3 -{k\over n}\rho_1\right),&\hbox{if}\ a\geq 2\cr}\right.\cr
& H^1(\Sigma_{rs}, f^*N_{C_{rs}/X})\simeq \left\{\matrix{
\oplus_{k=1}^{n-1} \left(\rho_2+{k\over n}\rho_1\right) \oplus 
\oplus_{k=1}^{n-1} \left(\rho_3+{k\over n}\rho_1\right), & \hbox{if}\ 
a=1\cr 
\oplus_{k=1}^{an-2}\left(\rho_2+{k\over n}\rho_1\right), \hfill 
& \hbox{if}\ a\geq 2.\cr}\right. 
\cr}}
Now we can finish the computation of the edge factors \edgefactorsC.\ 
We will consider the cases $a=1$ and $a\geq 2$ separately
\eqn\edgefactorsE{\eqalign{
& a=1:\qquad {e(H^1(\Sigma_{rs}, f_{rs}^*T_X)^m)\over 
e(H^0(\Sigma, f_{rs}^*T_X)^m)} = H^{-2}{\prod_{k=1}^{n-1}(n\rho_2+k\rho_1)
\prod_{k=1}^{n-1}(n\rho_3+k\rho_1)\over \rho_1^n (-\rho_1)^n ((n-1)!)^2}\cr
& a\geq 2: \qquad {e(H^1(\Sigma_{rs}, f_{rs}^*T_X)^m)\over 
e(H^0(\Sigma_{rs}, f_{rs}^*T_X)^m)} = H^{-2}{\prod_{k=1}^{an-2}(n\rho_2+k\rho_1)\over 
\rho_1^n (-\rho_1)^n ((n-1)!)^2 \prod_{n=0}^{n(a-2)}(n\rho_3-k\rho_1)}.\cr}}
Before comparing \edgefactorsB\ and \edgefactorsE\ one has to remember that 
the weights $\rho_i$, $i=1,2,3$ are supposed to satisfy the condition 
$\rho_1+\rho_2+\rho_3=0$ in order to preserve the lagrangian cycles. 
Using this condition, we can rewrite the expressions in \edgefactorsE\ 
as functions of $\rho_1, \rho_2$ only
\eqn\edgefactorsF{\eqalign{
& a=1:\qquad {e(H^1(\Sigma_{rs}, f_{rs}^*T_X)^m)\over 
e(H^0(\Sigma_{rs}, f_{rs}^*T_X)^m)} = (-1)^{n-1}H^{-2}
{\prod_{k=1}^{n-1}(n\rho_2+k\rho_1)
\prod_{k=1}^{n-1}(n\rho_2+(n-k)\rho_1)\over \rho_1^n (-\rho_1)^n ((n-1)!)^2}\cr
& a\geq 2: \qquad {e(H^1(\Sigma_{rs}, f_{rs}^*T_X)^m)\over 
e(H^0(\Sigma_{rs}, f_{rs}^*T_X)^m)} = (-1)^{1+n(a-2)}H^{-2}
{\prod_{k=1}^{n-1}(n\rho_2+k\rho_1)\prod_{k=1}^{n-1}(n\rho_2+(na-k)\rho_1)\over 
\rho_1^n (-\rho_1)^n ((n-1)!)^2}.\cr}}
Therefore we can conclude that 
\eqn\edgefactorsG{
{e(H^1(\Sigma_{rs}, f_{rs}^*T_X)^m)\over 
e(H^0(\Sigma_{rs}, f_{rs}^*T_X)^m)} = (-1)^{1+n(a-2)} 
{e(H^1(\Delta_r, \CT_{Xr})^m)\over 
e(H^0(\Delta_r, \CT_{Xr})^m)}
{e(H^1(\Delta_s, \CT_{Xs})^m)\over 
e(H^0(\Delta_s, \CT_{Xs})^m)}}
for all $a$. 
The last element we need is a similar formula for the contributions 
of the automorphism groups. One can easily check that 
\eqn\edgefactorsH{
e(\Aut(\Sigma_{rs})^m)= e(\Aut(\Delta_r)^m) e(\Aut(\Delta_s)^m).}

\listrefs
\end
An important element in the gluing algorithm is the automorphism group 
of a fixed open string map $(\Sigma,f)$. 
Since the boundary components of $\Sigma$ are ordered, the automorphism group 
is an extension 
\eqn\vpC{
1\ra \prod_{i=1}^3\prod_{a_i=1}^{h_i} (\IZ/{n^i_{a_i}}) \ra G \ra \prod_{i=1} \CS_{h_i} \ra 1.}
$\CS_{h_i}$ is the permutation group on $h_i$ letters acting on $f$ by permuting the 
components $\Delta^i_{a_i}$ of fixed $i$ preserving the winding numbers.
In the expression \vpA\ we sum over all ordered $h_i$-uples $n^i_{a_i}$.
Since the 
invariants $C(g,h|d_i, n^i_{a_i})$ are invariant under the action of $\CS_{h_i}$, 
we obtain the sum \vpB\ over winding vectors which parameterize unordered $h_i$-uples.
Each term in \vpB\  will be divided by an automorphism group factor of the form $\prod_{i=1}^3
\prod_{m=1}^{\infty} {1\over m^{k^i_m} k^i_m!}$.

Now we would like to find a gluing algorithm which generates all closed string graphs 
in terms of open string graphs. 
Let us first consider a particular edge in the graph 
$\Gamma$ corresponding to an invariant $(a,b)$ curve $C$ between the fixed points 
$P_L,P_R$. In the target space we have trivalent vertices centered at $P_L, P_R$ glued along 
a common pair of rays. To each of these vertices we can associate an open string generating 
functional of the form 
\eqn\vpD{
F_{L,R}(g_s,q_{L,R,i},V_{L,R,i}) = 
\sum_{\Lambda_{L,R}} g_s^{2g(\Lambda_{L,R})-2+h(\Lambda_{L,R})} C_{L,R}(\Lambda_{L,R}) 
\prod_{i=1}^3q_{L,R,i}^{l_i(\Lambda_{L,R})} \Upsilon_{|{\bf k}^i|(\Lambda_{L,R})}(V_{L,R,i})}
Each vertex consists of discs $D_{L,R,i}$ ending on lagrangian cycles $L_{L,R,i}$, $i=1,2,3$. 
Suppose $L_{L,3}=L_{R,3}$ are the same cycle dividing $C$ into the discs 
$D_{L,3}$ and $D_{R,3}$. 
The other discs $D_{L,2,3}$ and respectively $D_{R,2,3}$ are oriented along edges of 
$\Gamma$ which 
connect $P_L$ and $P_R$ to other fixed points on $X$. 

The first step in the algorithm is gluing the two local potentials \vpD.\

Localization gives the following formulae for the edge factors 
\eqn\vpF{\eqalign{ 
& F_r(e_r) = {e_T(H^1(\Delta_{r};f_r^*T_X, f_{r\partial}^*T_{L_{rs}})^m)e_T(\Aut(\Delta_{r}))^m
\over e_T(H^0(\Delta_{r};f_r^*T_X, f_{r\partial}^*T_{L_{rs}})^m)}\cr
& F_s(e_s) = {e_T(H^1(\Delta_{s};f_s^*T_X, f_{s\partial}^*T_{L_{rs}})^m)e_T(\Aut(\Delta_{s}))^m
\over e_T(H^0(\Delta_{s};f_s^*T_X, f_{s\partial}^*T_{L_{rs}})^m)}\cr
& F_{rs}(e_{rs})={e_T(H^1(\Sigma_{rs}, f_{rs}^*T_X)^m) e_T(\Aut(\Sigma_{rs})^m)\over 
e_T(H^0(\Sigma_{rs}, f_{rs}^*T_X)^m)}.\cr}}